\documentclass[useAMS,usenatbib]{mn2e}
\usepackage[latin1]{inputenc}
\usepackage{verbatim}
\usepackage{aas_macros}
\usepackage{graphicx}
\usepackage{amsmath}
\usepackage{color}

\def\lesssim {<\kern-1.2em\lower1.1ex\hbox{$\sim$}~} 

\title[Toward a consistent comparison between hydrodynamical simulations and SDSS]{Physical properties of galaxies: toward a consistent comparison between hydrodynamical simulations and SDSS }
\author[Guidi et al.]{Giovanni Guidi$^{1}$, Cecilia Scannapieco$^{1}$, Jakob Walcher$^{1}$ and Anna Gallazzi$^{2}$\\
$^1$ Leibniz-Institut f\"ur Astrophysik Potsdam (AIP), An der Sternwarte 16, D-14482, Potsdam, Germany\\
$^2$ INAF-Osservatorio Astrofisico di Arcetri, Largo Enrico Fermi 5, I-50125 Firenze, Italy
}

\begin{document}

\date{Accepted ... Received ...; in original form ...}

\pagerange{\pageref{firstpage}--\pageref{lastpage}} \pubyear{2014}

\maketitle

\begin{abstract}
We study the effects of applying observational techniques to 
derive the properties of simulated galaxies, with the aim of making 
an unbiased comparison between observations and simulations.
For our study, we used fifteen galaxies simulated in
a cosmological context using three different feedback and chemical
enrichment models, and compared their
$z=0$ properties  with data from the Sloan Digital Sky Survey (SDSS). 
We show that the physical properties obtained directly from the
simulations without post-processing can be very different
to those obtained mimicking observational techniques. 
In order
to provide simulators a way to reliably compare their galaxies with SDSS data,
for each physical property that we studied -- colours, magnitudes,
gas and stellar metallicities, mean stellar ages
and star formation rates -- we give scaling relations that can be easily
applied to the values extracted from the simulations; 
these scalings have in general a high correlation, except for
the gas oxygen metallicities.
Our simulated galaxies are photometrically  
similar to galaxies in the blue sequence/green valley, 
but in general they appear older, passive and with lower metal content
compared to most of the spirals in SDSS.
As a careful assessment of the agreement/disagreement with observations 
is the primary test of the baryonic physics implemented in 
hydrodynamical codes, our study shows that considering the observational 
biases in the derivation of the galaxies' properties is of fundamental
importance to decide on the failure/success of a galaxy formation model.
\end{abstract}


\section{Introduction}

The various
physical processes that occur at  galactic scales
during galaxy evolution 
leave imprints on the shape and features of the galaxies' Spectral
Energy Distributions (SEDs), which represent the main 
source of knowledge about the properties of galaxies in the 
Universe. Observational algorithms are able to recover, from a galaxy's SED,
its physical properties  such as the stellar, 
gas and 
metal content, the conditions of the InterStellar Medium  (ISM) and 
the properties of Active Galactic Nuclei (AGNs). 
These methods can provide constraints on different physical properties, 
either exploiting a pixel-by-pixel fit to the spectrum (e.g.
\citealt{Cappellari04};  \citealt{CidFernandes05}; \citealt{Ocvirk06}; 
\citealt{Walcher15}) or interpreting single or few spectral 
features, for instance the emission line luminosity 
\citep{Kennicutt98}, emission line ratios 
\citep{Kobulnicky04}, Lick absorption indices \citep{Gallazzi05} 
or the  $4000 ~\rm{\AA{}}  $ break \citep{Bruzual83}.

With the advent of large surveys such as 
2dFGRS, SDSS,  2MASS, ALMA, HUDF, DEEP2, SPITZER, HERSCHEL
\citep{Colles99, Abazajian03, Werner04,  Beckwith06, Skrutskie06, Pilbratt10, Hodge13, Newman13}, 
huge datasets of galaxy photometric and
spectral informations at different 
wavelengths and redshifts are now available to observers. 
 Moreover, Integral Field Unit 
(IFU) spectrographs (e.g. MUSE, \citealt{Bacon04}; WEAVE, 
\citealt{Dalton14}) are opening the possibility to study spatially-resolved 
properties of nearby galaxies, and thanks to IFU surveys
such as CALIFA \citep{Sanchez12,Garcia-Benito15}, MANGA \citep{Bundy15} and 
SAMI \citep{Allen15}, two-dimensional spectral maps of galaxies are 
now available, providing in turn a more 
comprehensive view of a galaxy's formation and evolution.
The analysis of these large datasets enables astronomers to 
study the properties of galaxies in the Universe 
at different redshifts, and has allowed the derivation of important 
global quantities and relations, 
such as the Cosmic Star Formation History \citep{Madau14}, the Stellar 
Mass function \citep{Baldry12, Song15} and the Mass-Metallicity 
relation \citep{Tremonti04, Erb06}. 

This wealth of data is also useful to test the results 
of hydrodynamical simulations of galaxy formation in 
cosmological context, which 
follow the linked evolution of gas and dark matter from the early stages
of collapse (assuming  recipes for star formation, Supernovae 
and AGN feedback and chemical enrichment) and are able to connect the observed 
galaxy properties with a galaxy's  merger and accretion history 
 \citep[e.g.,][]{Scannapieco09, Oppenheimer10, Naab14, Nuza14, Vogelsberger14, 
Creasey15, Nelson15, Scannapieco15, Schaye15}. 
The results of hydrodynamical simulations have also been
used recently to test and calibrate observational diagnostics, as they 
have the advantage
that, unlike in observations,  the 
physical properties of the galaxies are known a priori
\citep{Belovary14, Michalowski14, Hayward14, Smith15, Hayward15}.
However, since the simulations contain information on the mass
distributions, while observations are based on the analysis of 
the light emitted 
by  galaxies, in general the various physical properties are
differently defined in observational and simulation studies.
For this reason,  a simple comparison
between them might not be reliable \citep[e.g.,][]{Scannapieco10},
and a suitable conversion of simulations into mock observations is required
in order to perform an unbiased comparison.
The creation of mock observations of simulated galaxies
is possible using different approaches, including
radiative transfer computations that follow the generation and propagation 
of light in a dusty ISM (e.g. \citealt{Jonsson10, 
Dominguez-Tenreiro14}).

In order to derive the galaxy properties from the mock SEDs,
it is also important to take into account that
each galaxy survey suffers from
observational biases related to the
 observational setup and strategy, as 
well as from the different assumptions 
in the pipelines used to derive the physical properties \citep{Walcher11}.
The existence of these observational effects can
have a strong influence in
 the comparison between simulations and observations, as it has 
already been shown that various methods to derive galaxy
properties show large variations
\citep{Scannapieco10,Michalowski14, Hayward15, Smith15}.

In this paper we study the effects of properly taking into
account the observational biases when 
simulations are compared with data from the Sloan Digital Sky 
Survey (SDSS). For this, we mimic the derivation of galaxy
properties done in SDSS, and consider the biases of this survey. 
In the companion paper \citealt{Guidi15} (hereafter PaperI) we discussed 
the models used to generate mock SEDs from fifteen simulations that
we use in our study, and we 
compared the results of applying different observational estimators on the
values derived for the galaxies' colours, magnitudes, stellar masses,
star formation rates, gas and stellar metallicities and mean stellar
ages. 
 In particular, we focused on the ability of the observational methods 
to recover galaxy properties close to the ones calculated 
directly from the simulations without post-processing,  
and we made a detailed study
on the biases and systematics of the different estimators. 

We found in PaperI that biases and systematics
affect all galaxy properties at different levels, and arise mainly 
from the specific design of the  observational setup,
from using
mass-weighted or luminosity-weighted averaged quantities, from the 
different parametrization of the template of models fitted to the 
spectrum/photometry of a galaxy, and from the calibrations assumed to derive
the gas metallicity and star formation rate (SFR).
These results are used in this work, 
in which we focus on an unbiased comparison
between the simulated galaxies and observations of SDSS.

The structure of the paper is as follows. In Section 2 we describe the 
simulations and the feedback and chemical models used to generate the galaxy sample,
as well as the methods used to create the synthetic observations.
In Section 3 we present the observational dataset and we explain our
selection of a subsample of type-classified galaxies. In Section 
4 we compare the different physical properties of the simulated galaxies with 
SDSS, and we provide fitting functions to convert the 
values derived directly from the simulations into the ones extracted
observationally.
Finally, in Section 5, we give our conclusions.


\section{Methodology}

\subsection{The simulations}

We use in this work
cosmological hydrodynamical simulations of galaxy formation
that are based on the dark-matter only 
Aquarius simulations \citep{Springel08}. 
In particular, we use five galaxy halos 
with present-day virial mass similar to the Milky
Way, i.e. 
$ 0.7 \times 10^{12}$M$_\odot < M_{200} < 1.7\times 10^{12}$M$_\odot$
(see \citealt{Scannapieco09} for 
details).  
Each of the five halos is then re-simulated
up to the present time
including a baryonic component using the zoom-in 
technique \citep{Tormen97}, with three different hydrodynamical codes 
based on Gadget-3 
\citep{Springel05} which assume various recipes for 
star formation, chemical enrichment, metal-dependent cooling and 
SNe feedback, composing a total of fifteen galaxies. 
As shown in \citet{Scannapieco12}, differences in the implementation of 
SNe feedback have strong effects on the properties of  simulated 
galaxies, and different hydrodynamical codes can produce
galaxies with a large range of physical properties (e.g. morphologies, 
sizes, metallicities, ages, star formation rates) even for the same 
dark-matter halo.

In order to identify the different galaxies we assign letters from A to E
for the five dark-matter halos, adding a label for the hydrodynamical code with
which the galaxies have been simulated:  either CS, 
CS$^+$ or MA.
Simulations CS
are run with the model 
described in \citet{Scannapieco05,Scannapieco06}, 
which includes star formation, chemical enrichment, metal-dependent
cooling, feedback from supernova Type Ia and TypeII, and a multiphase
model for the ISM.
The second set of five simulations labelled CS$^+$ is generated with 
an updated version of the \citet{Scannapieco05,Scannapieco06} model
by Poulhazan et al. (in prep.), which adopts different choices
for the  chemical yields, a Chabrier Initial Mass Function (IMF),
and includes chemical feedback from AGB stars.
The third set, referred to as MA, is simulated with 
the update to the Scannapieco et al. code
by \cite{Aumer13}; the main changes are a different set of chemical
yields (which also include AGB stars), a different metal-dependent cooling 
function and a Kroupa IMF. In addition, the code has a different 
implementation of energy feedback from SNe, which is divided 
into a thermal and a kinetic part, and it includes feedback 
from radiation pressure coming from massive young stars. The
MA 
model has in general stronger feedback compared to the CS/CS$^+$ models, 
resulting in more disky,  younger galaxies.
All the fifteen galaxies of our sample have, at redshift $z=0$, 
total stellar masses between 
$1-10\times 10^{10}$M$_\odot$, 
gas masses in the range $3-10\times 10^{10}$M$_\odot$, 
stellar/gas mass resolution of $2-5\times 10^{5}$M$_\odot$, 
dark matter particle mass of $1-2\times 10^{6}$M$_\odot$,
and gravitational softening of $300-700$ pc. The
cosmological parameters assumed are:
$\Omega_{\rm m} = 0.25 $, $\Omega_{\Lambda} = 0.75$,  $\Omega_{\rm b} = 0.04$, 
$\sigma_8 = 0.9$ and $H_0 = 100 \, h$~km~s$^{-1}$~Mpc$^{-1}$ with $h=0.73$.

\subsection{Creating the mock observations}

The hydrodynamical simulations at redshift $z=0$ have 
been post-processed with the 
radiative transfer code {\sc sunrise} \citep{Jonsson06, Jonsson10},
which simulates the propagation of light through
a dusty ISM using Monte Carlo techniques, and 
self-consistently derives the spectra of the simulated galaxies from
different observing positions, including stellar/nebular emission, 
dust absorption and IR-emission.
In a first stage, {\sc sunrise} assigns each star particle a spectrum.
For star particles older than 10 Myr, the stellar spectrum is selected 
according to the age, metallicity and mass of the particle from a template
of spectra generated with the stellar population synthesis code
{\sc starburst99}, 
choosing the Padova 1994 stellar tracks \citep{Fagotto94,Fagotto94_1}, 
a Kroupa IMF \citep{Kroupa02} (with $\alpha = 1.3$ for 
$m_{\text{star}} = 0.1 - 0.5 \; \text{M}_{\odot}$ and $\alpha = 2.3$ for 
$m_{\text{star}} = 0.5-100 \; \text{M}_{\odot}$) and Pauldrach/Hillier stellar 
atmospheres.
On the other hand, for star particles  younger than 10 Myr, {\sc sunrise} assigns a 
nebular spectrum that takes into account the effects of photo-dissociation and 
recombination of the surrounding gas.
The nebular spectra are pre-computed with 
the photo-ionization code {\sc mappings III} \citep{Groves04,Groves08}, and 
depend on the metallicity of the star particle and the gas 
around it, on the ISM pressure\footnote{The ISM pressure enters
the {\sc mappings} computation through the compactness parameter $C$, which is also 
related to the assumed cluster mass M$_{cl}$ (see  
\citealt{Groves08, Jonsson10}).}, and on the Photo-Dissociation Region (PDR)
covering fraction $f_{\text{PDR}}$.
The {\sc mappings III} parameters not constrained by the underlying 
hydrodynamical simulation,  $f_{\text{PDR}}$ and M$_{cl}$, have been set 
respectively to $f_{\text{PDR}} = 0.2$ 
and M$_{cl} = 10^5 \, M_{\odot}$, following \citet{Jonsson10}.

Once  a spectrum is assigned to each star particle, {\sc sunrise} 
enters the radiative 
transfer stage, where random-generated photon ``packets'' (rays) 
are propagated from 
these sources through the ISM using a Monte Carlo approach (we use 
$\sim 10^7$ Monte Carlo rays in our case\footnote{We tested the 
effects of increasing $10$ times the number of rays, and found
very small relative differences on the SEDs, the largest
ones of the order of  $0.2-0.3\%$ occurring for the
edge-on projections, see appendix B.}). It is assumed 
that the dust is traced by the metals with a constant dust-to-metals 
ratio of 0.4 \citep{Dwek98}, and that dust extinction is 
described by a Milky Way-like extinction curve normalized to $R_V =3.1$ 
\citep{Cardelli89,Draine03}. In our model, unlike in other hydrodynamical 
codes where gas particles represent a mix of gas/stellar phases, (e.g.
\citealt{SH03}), each gas particle has a single temperature, density and 
entropy (see \citealt{Scannapieco06} for details), and so the amount of dust 
is linked to the total amount of metals in the gas particle 
(see also \citealt{Hayward11, 
Snyder13, Lanz14}).

The tracing of the rays is done on an adaptive grid, which for our simulations 
is represented by a number of cells between 
$\sim 30.000 - 400.000$\footnote{We have also run {\sc sunrise} increasing
the number of cells
$\sim 10-20$ times, and found very small relative differences
in the derived spectra, the largest ones occurring in the case of the
edge-on SEDs, and being of the order  of $0.5-1\%$, see appendix B.}, 
and covers a box with side 120 kpc, with minimum cell size 
of $\sim 220-460$ pc.
To compute the grid, we have assumed a value of tolerance
tol$_{met} = 0.1$ and metals opacity  
$ \kappa = 3 \times 10^{-5}$ kpc$^2$ M$_{\odot}^{-1}$ following \citet{Jonsson06}.

We have also followed a simple approach by  applying 
Stellar Population Synthesis (SPS) models to derive
some of the galaxy properties (see PaperI), which we discuss
here when appropriate.

\begin{figure*}
  \centering
  {\textbf{A-CS\hspace{2.cm}B-CS\hspace{2.cm}C-CS\hspace{2.cm}D-CS\hspace{2.cm}E-CS}\par\medskip\vspace{-0.1cm}\includegraphics[width=2.7cm]{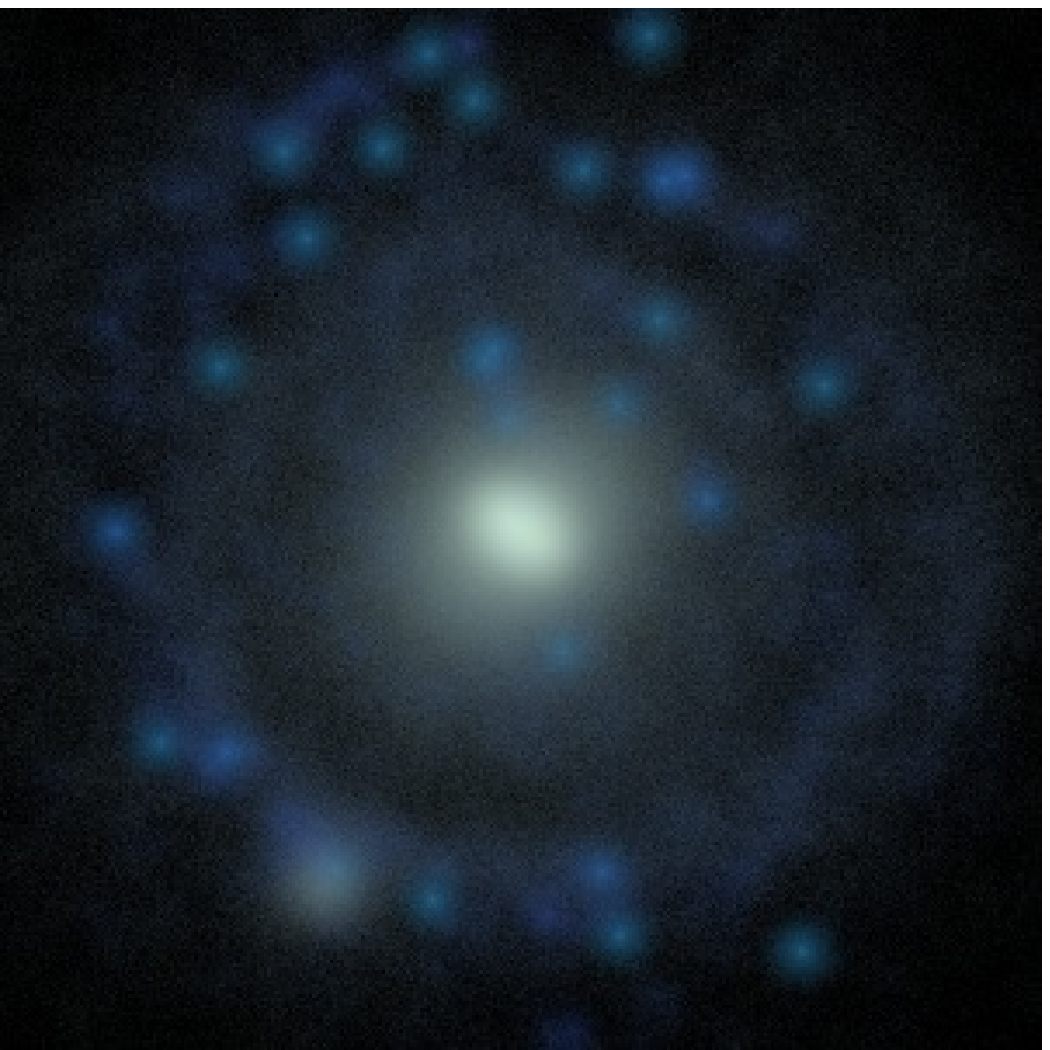}}\hspace{0.1cm}{\includegraphics[width=2.7cm]{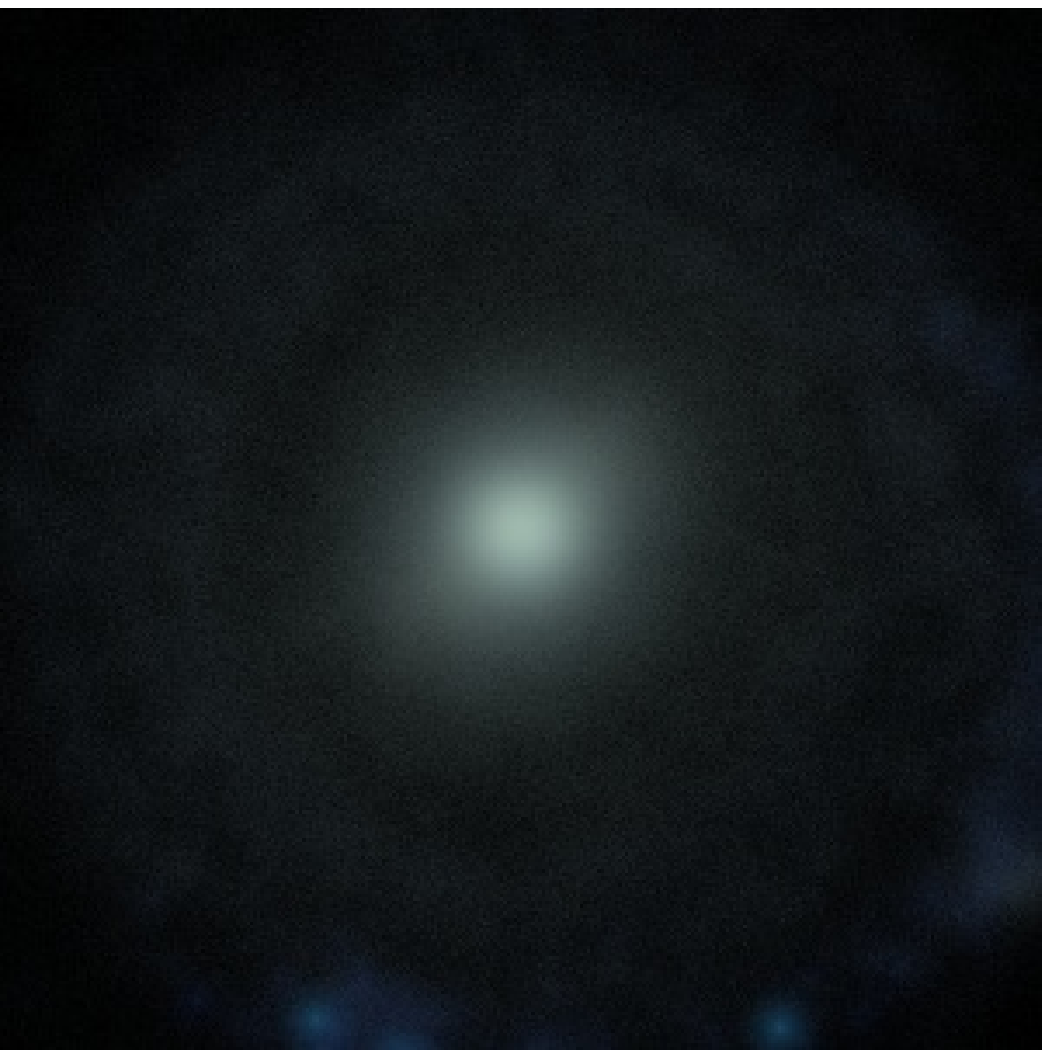}\hspace{0.1cm}\includegraphics[width=2.7cm]{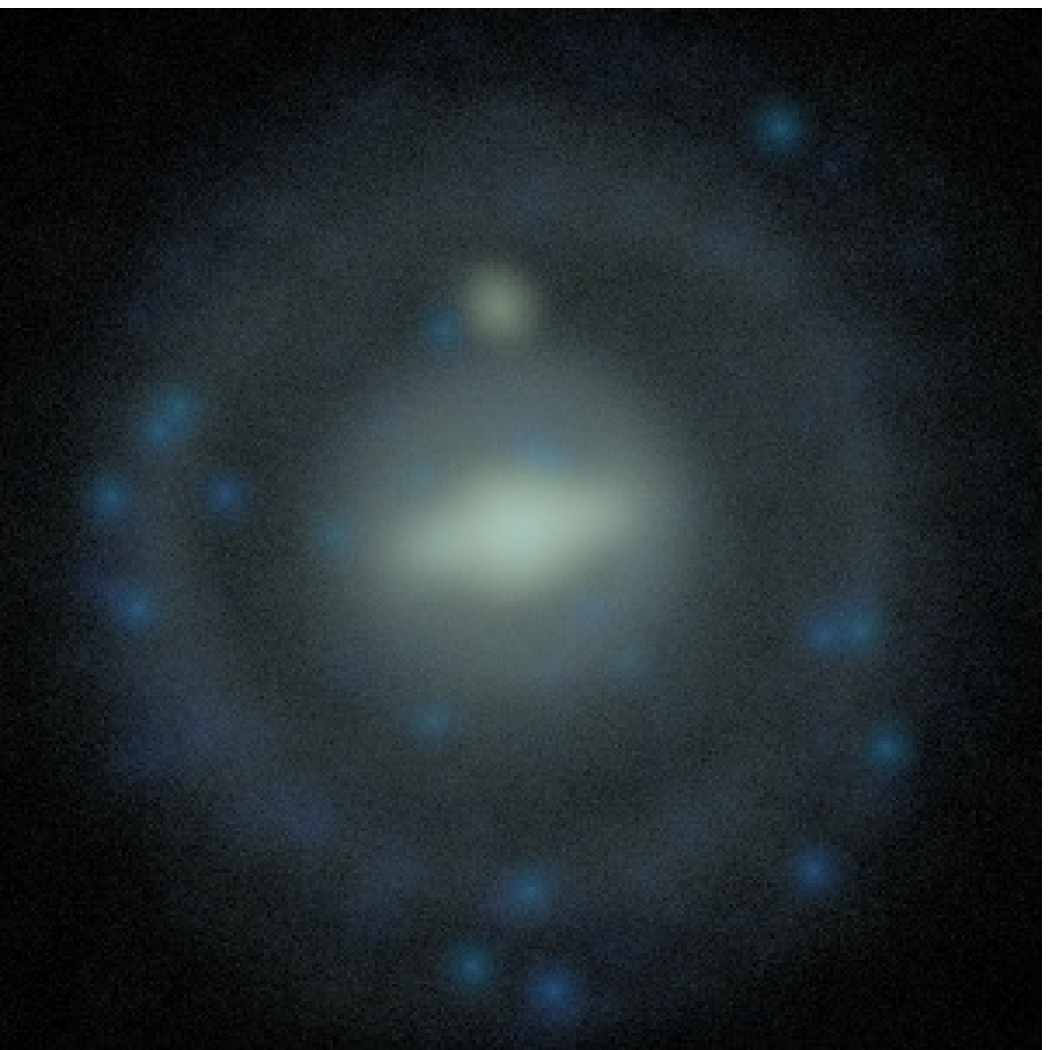}\hspace{0.1cm}\includegraphics[width=2.7cm]{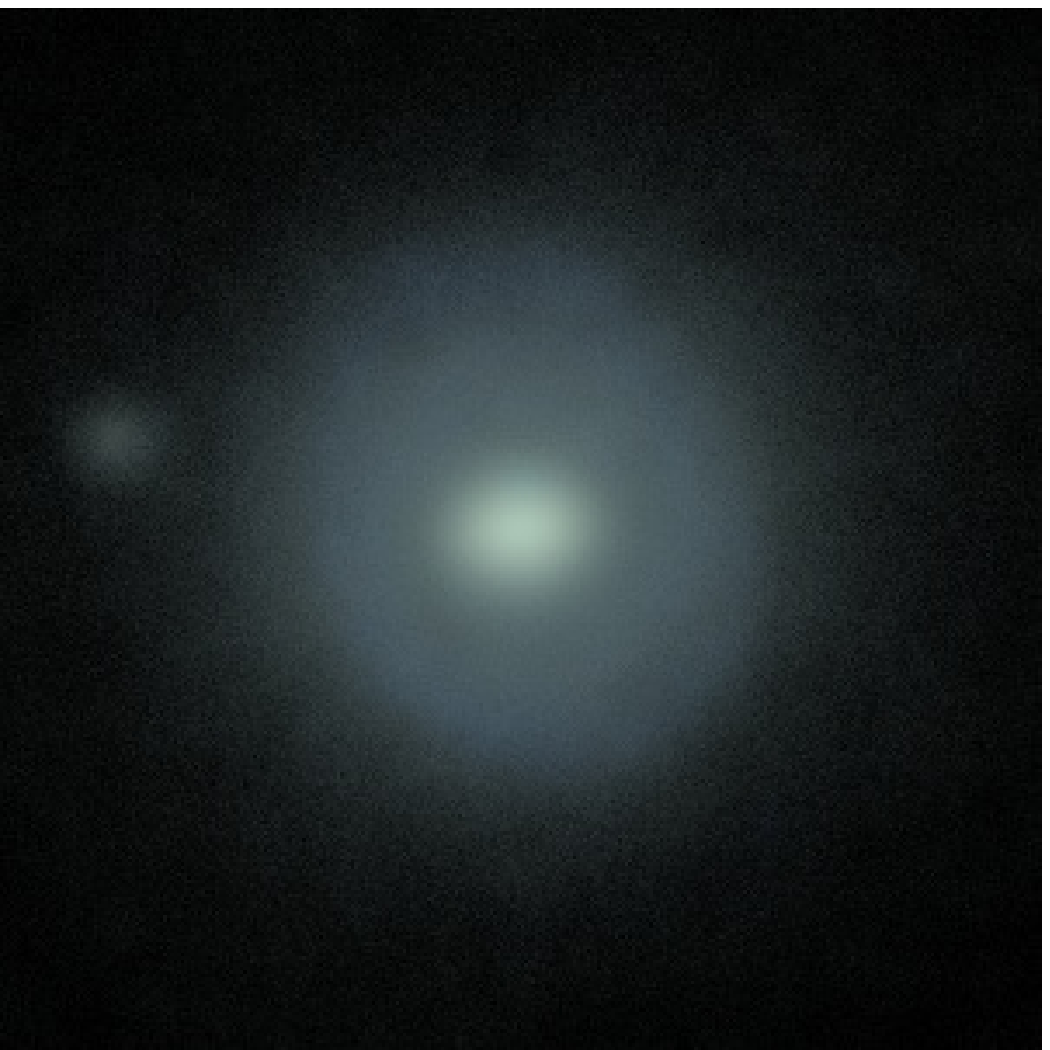}\hspace{0.1cm}\includegraphics[width=2.7cm]{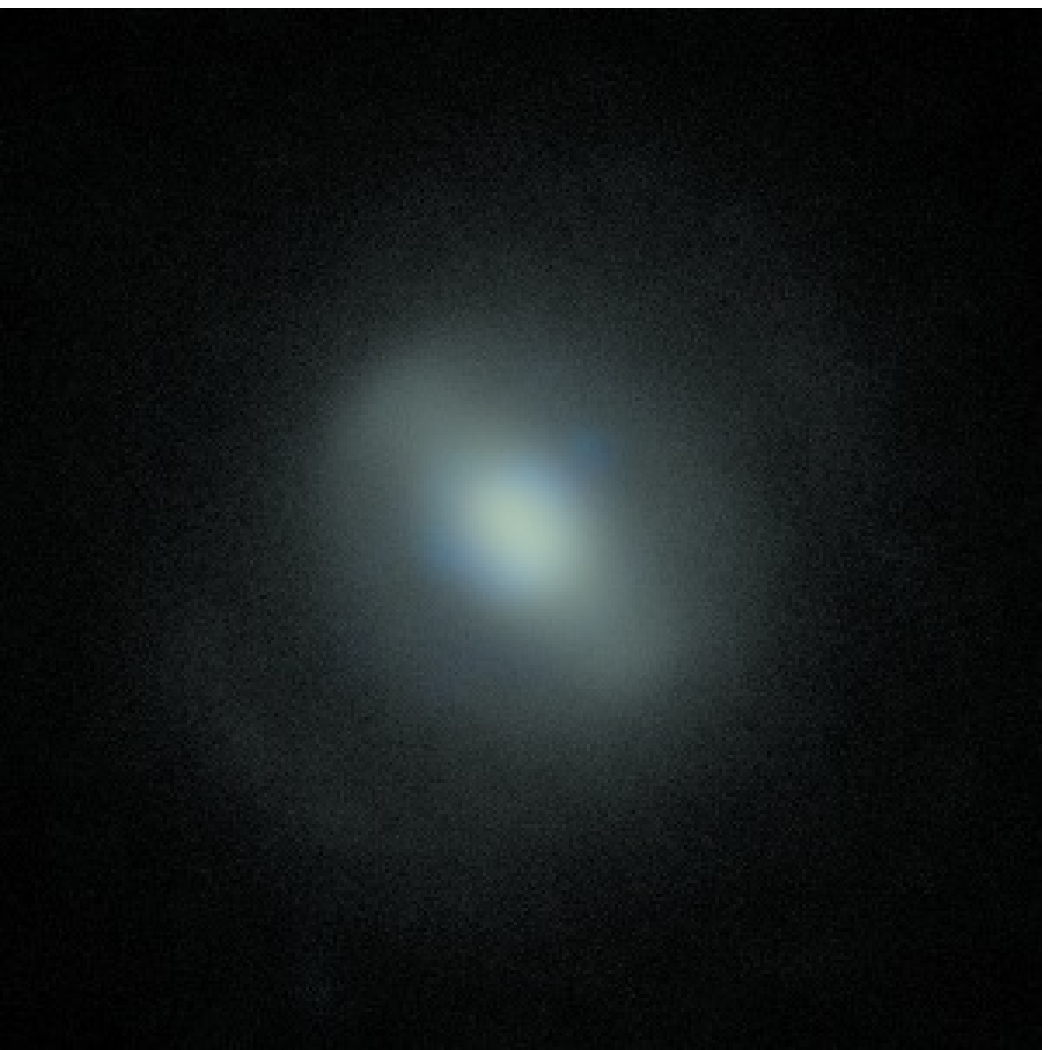}}

\vspace{0.2cm}

  {\textbf{A-CS$^+$\hspace{1.8cm}B-CS$^+$\hspace{1.8cm}C-CS$^+$\hspace{1.8cm}D-CS$^+$\hspace{1.8cm}E-CS$^+$}\par\medskip\vspace{-0.1cm}\includegraphics[width=2.7cm]{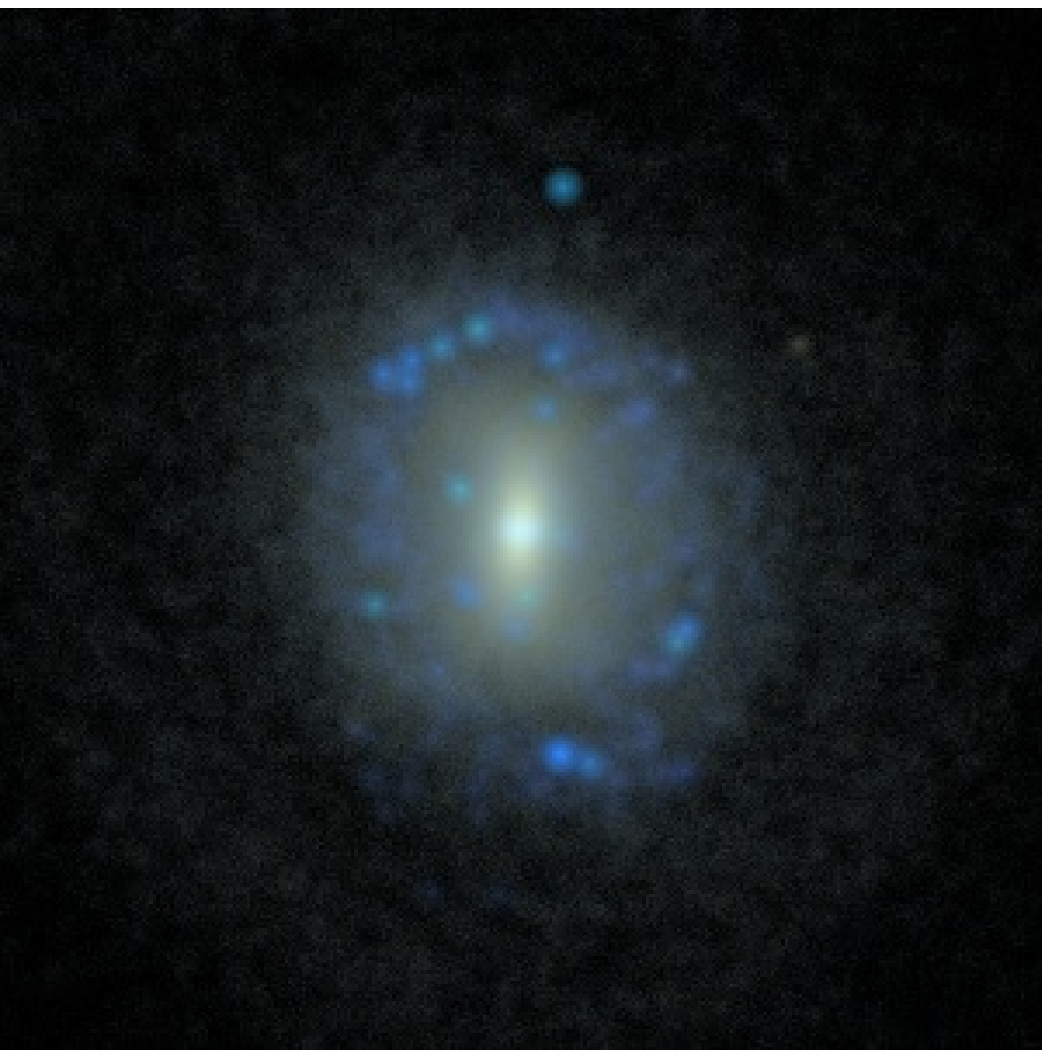}}\hspace{0.1cm}{\includegraphics[width=2.7cm]{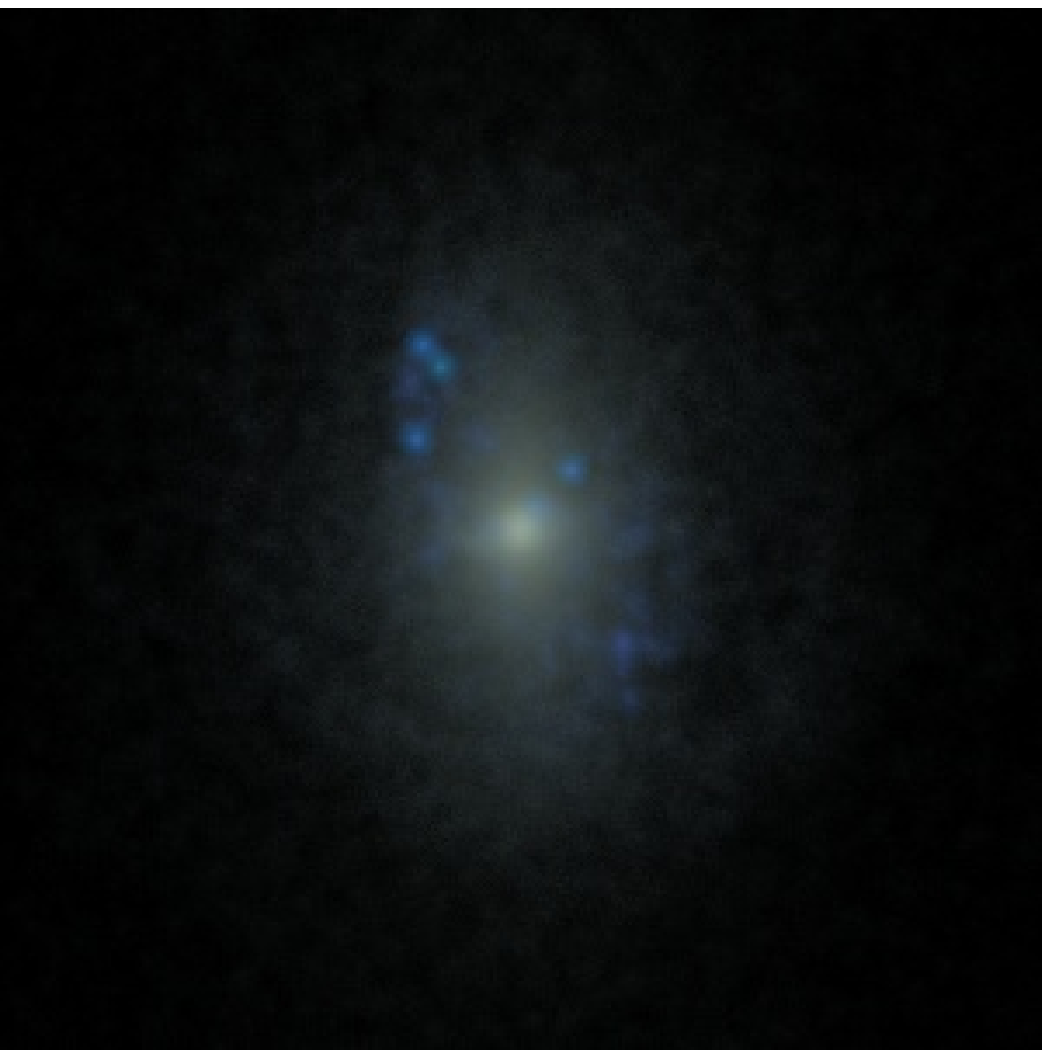}\hspace{0.1cm}\includegraphics[width=2.7cm]{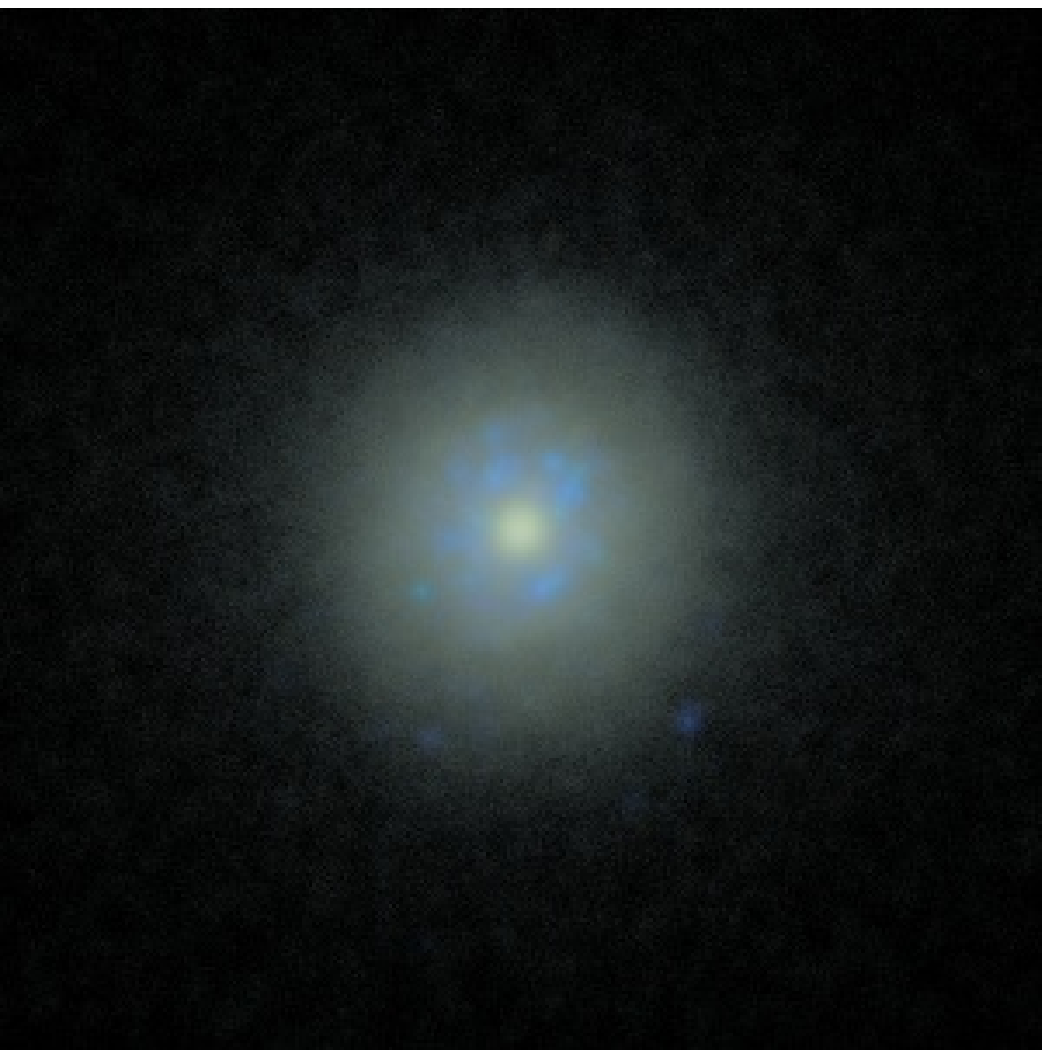}\hspace{0.1cm}\includegraphics[width=2.7cm]{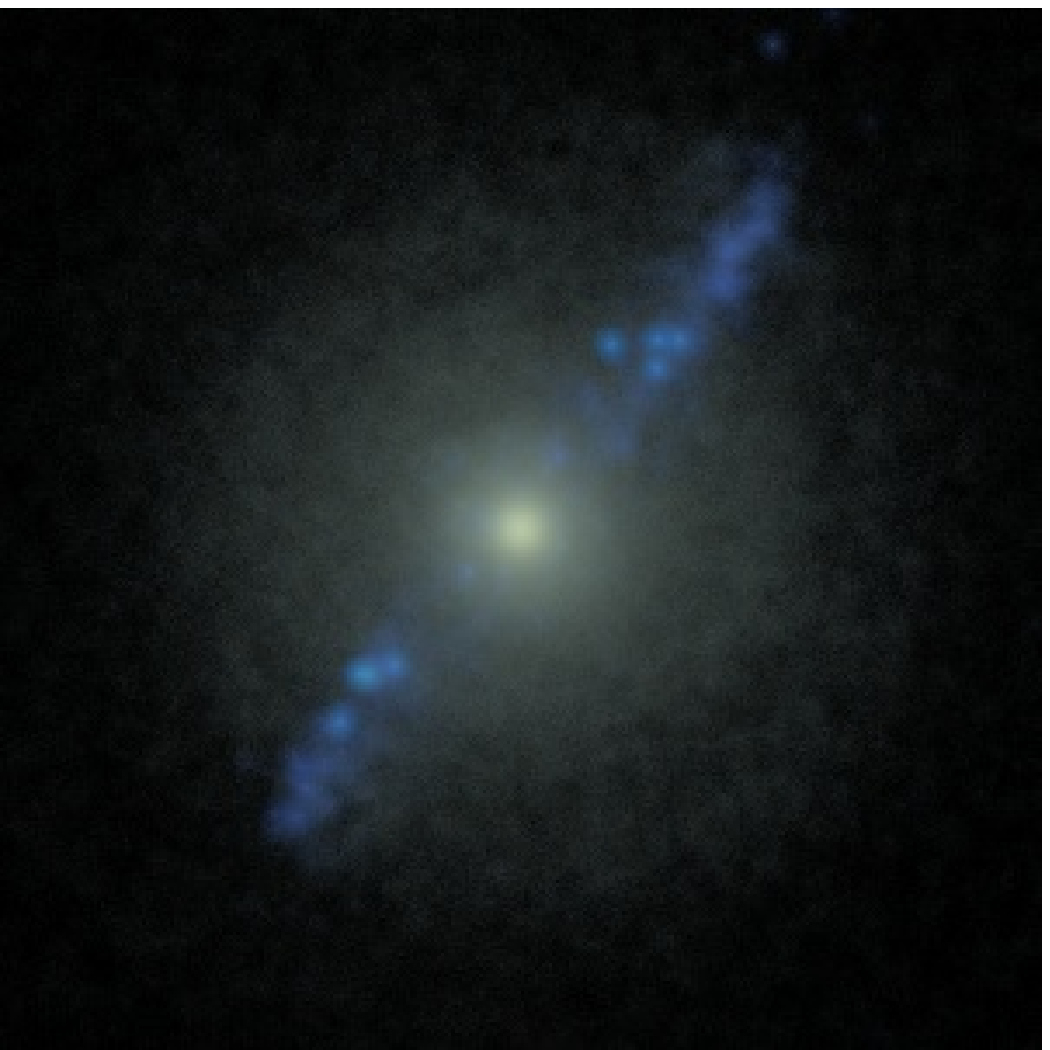}\hspace{0.1cm}\includegraphics[width=2.7cm]{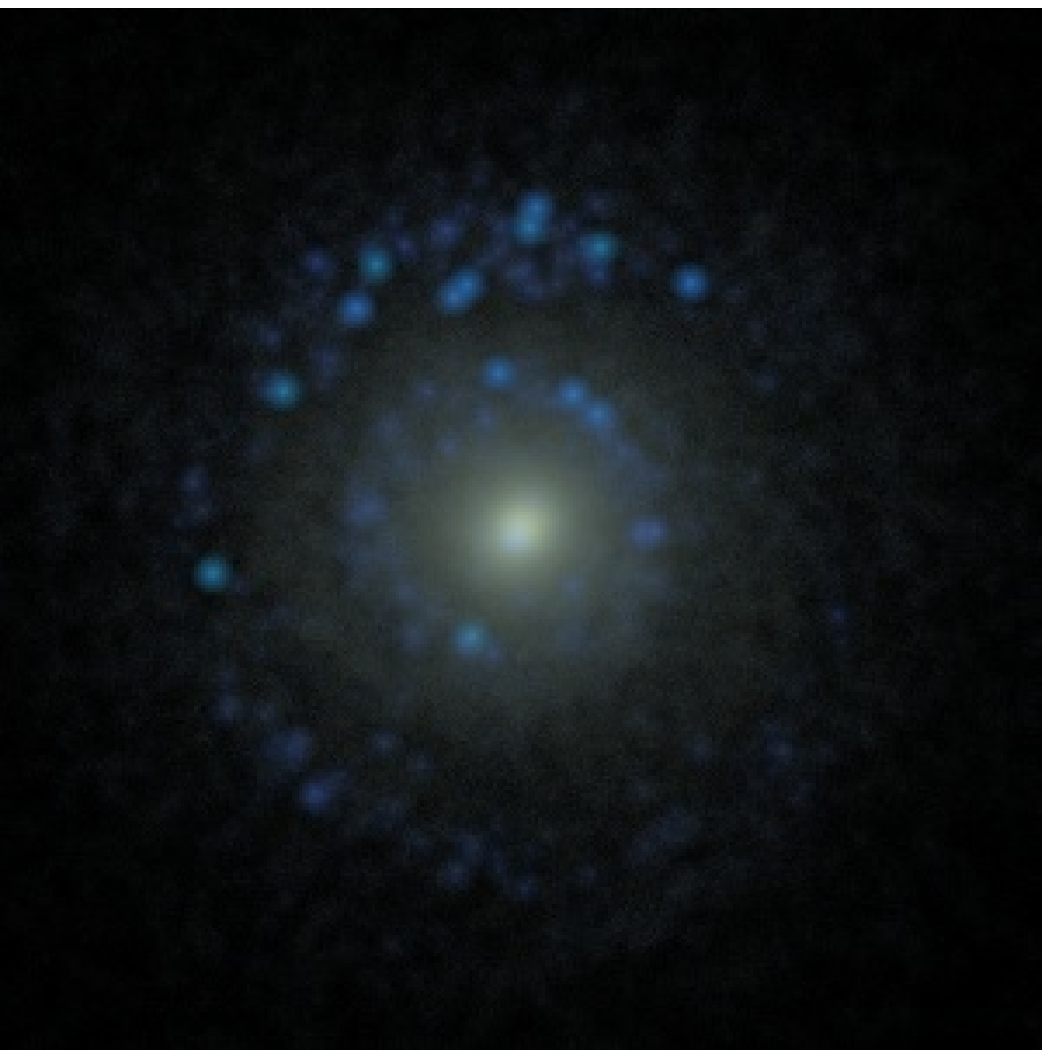}}

\vspace{0.2cm}

  {\textbf{A-MA\hspace{1.9cm}B-MA\hspace{1.9cm}C-MA\hspace{1.9cm}D-MA\hspace{1.9cm}E-MA}\par\medskip\vspace{-0.1cm}\includegraphics[width=2.7cm]{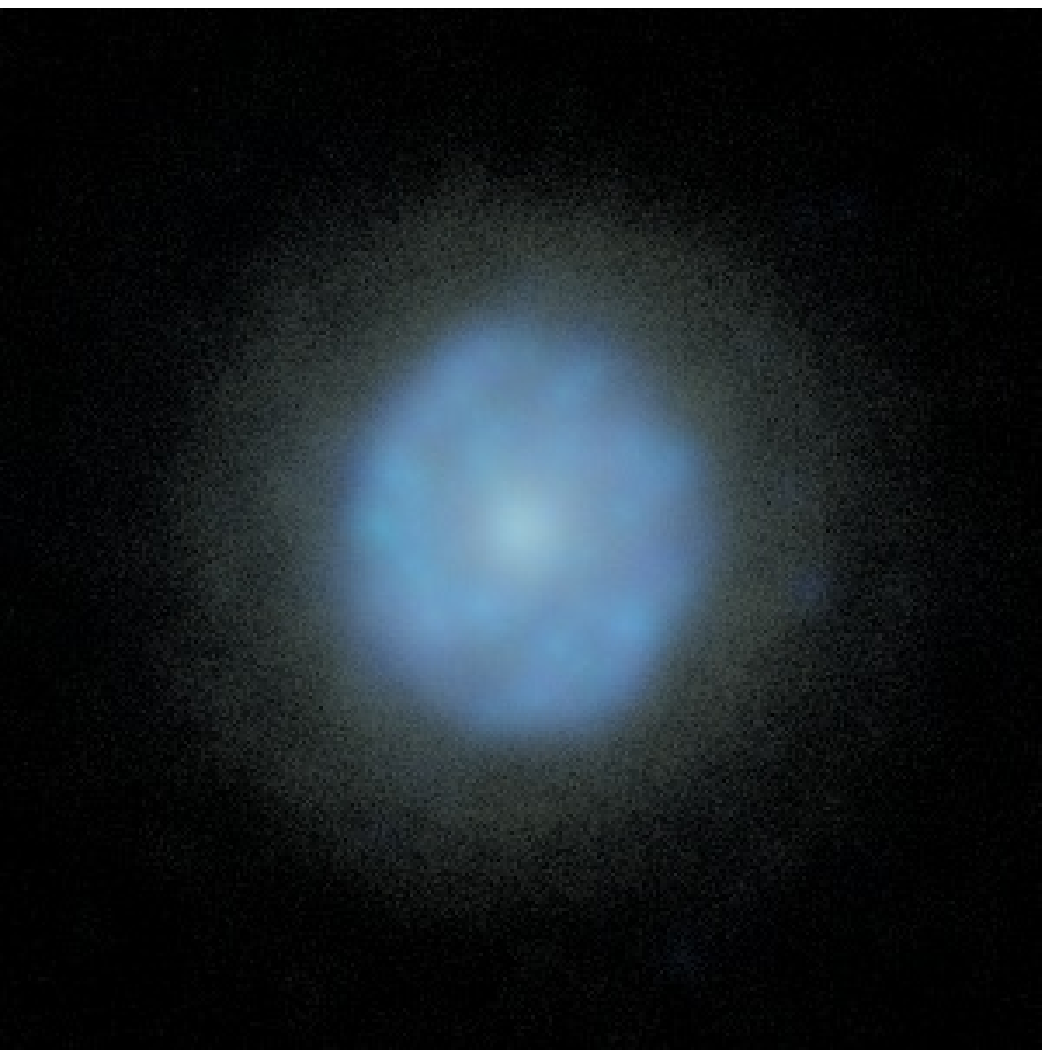}}\hspace{0.1cm}{\includegraphics[width=2.7cm]{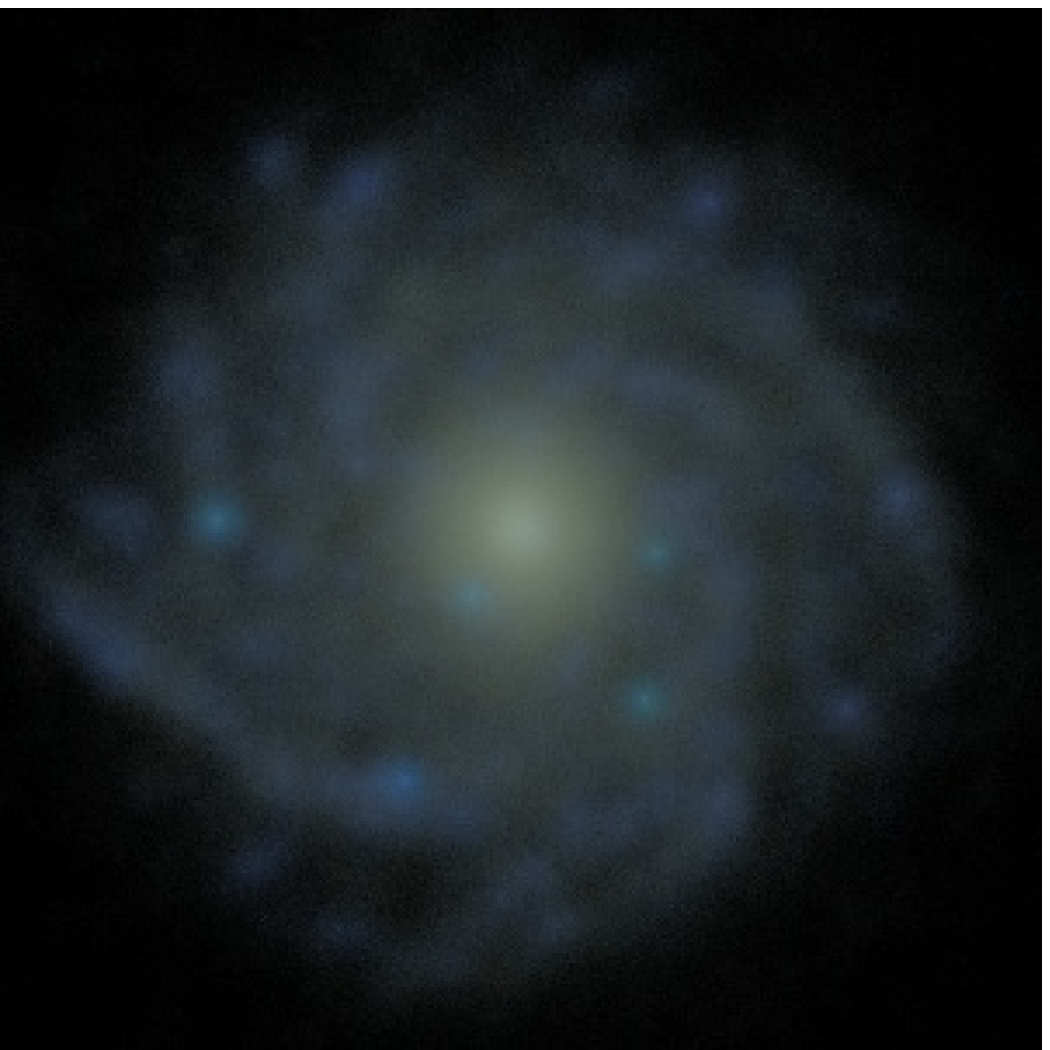}\hspace{0.1cm}\includegraphics[width=2.7cm]{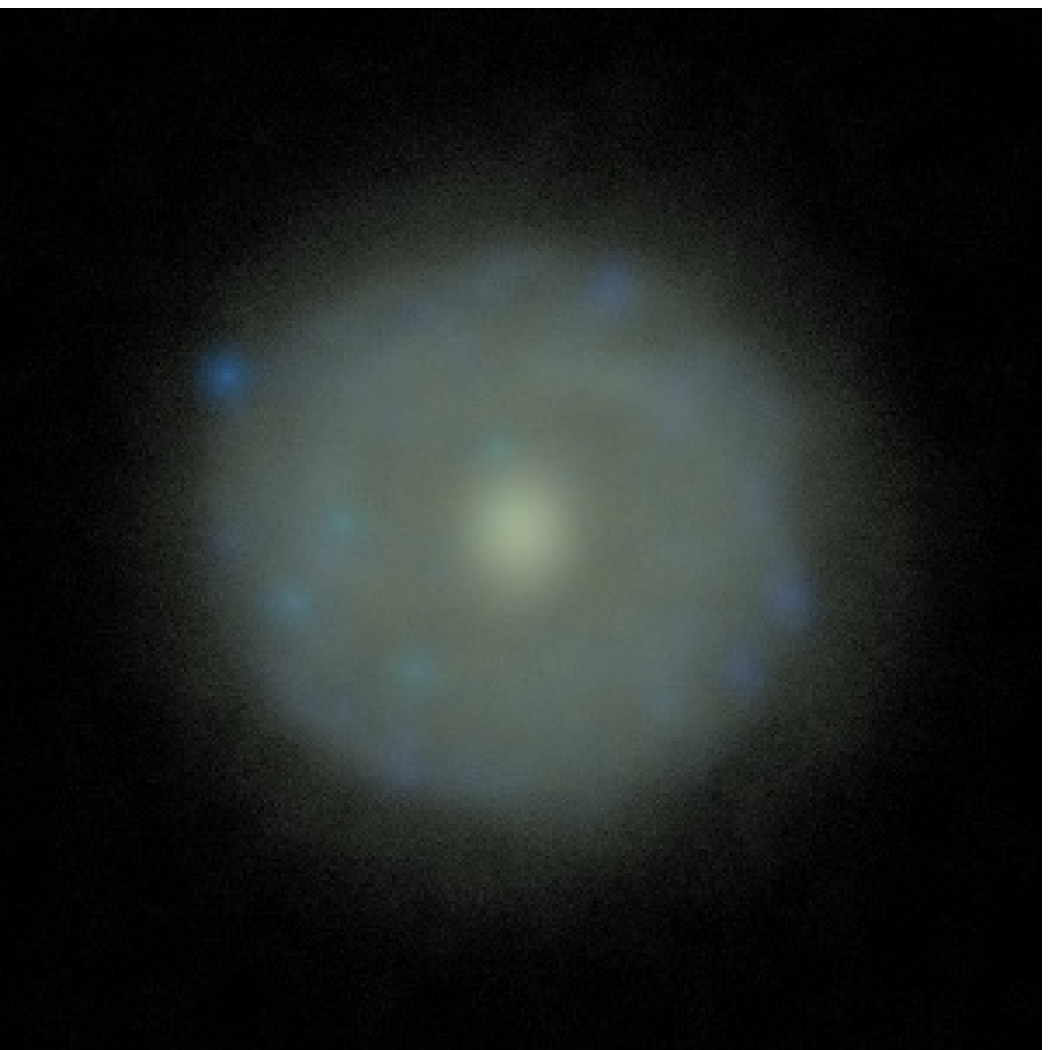}\hspace{0.1cm}\includegraphics[width=2.7cm]{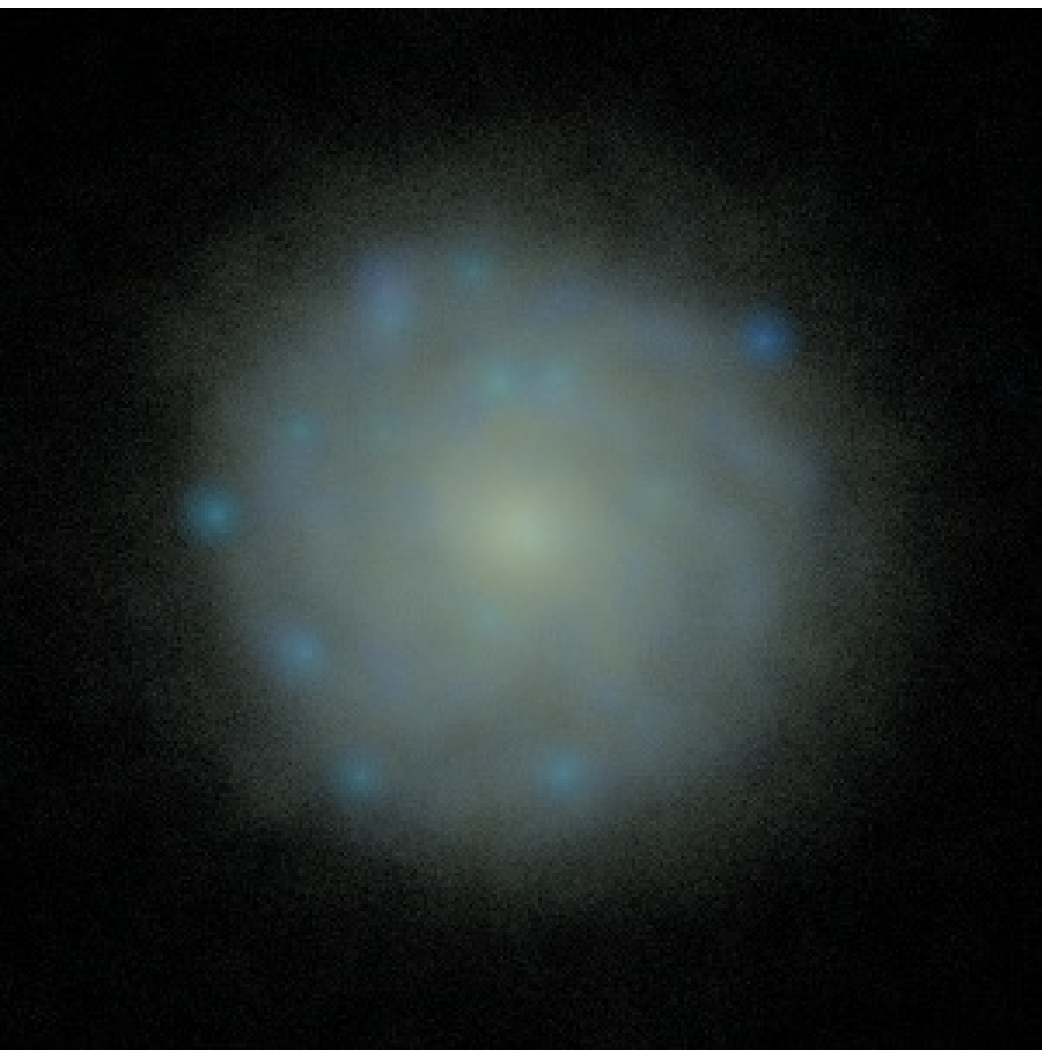}\hspace{0.1cm}\includegraphics[width=2.7cm]{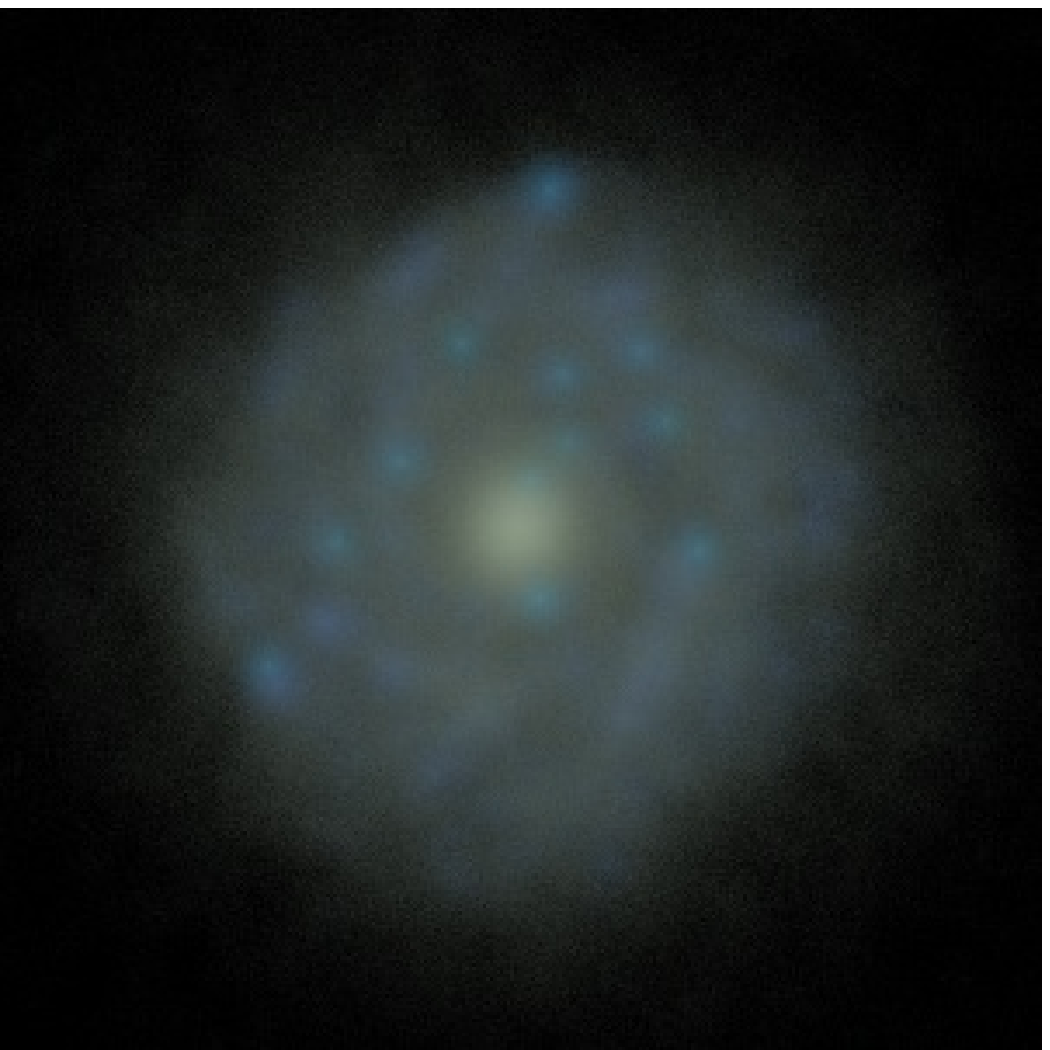}}

\vspace{0.2cm}

\caption{$(u,r,z)$ multi-band face-on images of the fifteen simulated galaxies, as predicted from the radiative transfer calculation of {\sc sunrise}, for a 60x60 kpc Field of View.}
\label{fig:galaxy_images}
\end{figure*}


\section{Observational data}
\label{sec:observations}

We compare the properties of our simulated galaxies, at 
redshift zero, with the Sloan Digital Sky Survey (SDSS) dataset. The SDSS 
camera \citep{Abazajian03} is designed to collect 
multi-color images and spectra of a large number of objects in an area of a 
third of the sky, at median redshift (for galaxies) of $z \sim 0.1$.  
The images are taken in 5 different photometric bands (\textit{u, g, r, i, z}) 
of increasing effective wavelength, with filter curves defined for an 
airmass of 1.3 at the Apache Point Observatory, pixel size of 0.396'' and 
exposure time of 53.9 s (\citealt{Gunn98}, \citealt{Gunn06}).

SDSS magnitudes are based on the AB photometric system \citep{Oke65, Oke83},
which allows immediate conversion from magnitudes to physical 
fluxes \citep{Fukugita96}.
The spectrographic survey observes spectra of 
$\sim 640$ target objects simultaneously, and the light of each object is 
collected with a single optical fibre 
of diameter 3 arcsec in the sky pointing at the center of the object
(see \citealt{York00,Smee13} for a technical description, see also PaperI
and \citealt{Stevens14} for a discussion of 
the effects of the limited fibre size on 
simulation properties). 
The wavelength covering is between 3800 and $9200 \rm{\AA{}} $ 
at resolution $R = 1800-2200$, and $\text{S/N} > 4$ at $g$-mag = 20.2.
In a PLANCK cosmology \citep{Planck15}, at $z \sim 0.1$, 
the fibre encloses a circular region of $\sim 5.8$ kpc in diameter, 
sampling only $\approx 1/3$ of the total light for a typical spiral 
galaxy \citep{Brinchmann04}. The SDSS Data Release 4 
(DR4, \citealt{Adelman06}) includes an imaging catalogue of about
180 million objects and spectroscopic data of $\sim 850.000$ objects,
of which $\sim 565.000$ are galaxies. 
The DR7 \citep{Abazajian09} contains photometric information of more 
than 350 million objects and spectra of $\sim 930.000$ galaxies.

In this work, we use datasets of derived galaxy properties from 
the MPA-JHU analysis of SDSS DR4 (for stellar masses/ages/metallicities) and 
DR7 (magnitudes/colors, stellar masses, gas metallicities, SFRs). 
From the MPA-JHU datasets we first select galaxies in 
the local universe ($z < 0.3$), 
and we separate the sample into early and late types 
according to visual classification.
For the DR4 data we use the \citet{Nair10} catalogue, which includes 
$\sim 14.000$ galaxies, while for DR7 we select 
early/late-type galaxies according to the Galaxy Zoo classification 
\citep{Lintott08, Lintott11}, a crowdsourcing-based project 
of morphological classification of $\approx 1/3$  of the SDSS DR7 
galaxies. 
In order to better estimate how close/far from real spirals 
our simulated galaxies are, 
we further split the samples (both for DR4 and DR7) in 
{\it green valley} galaxies \citep{Martin07} 
defined according to the \citet{Salim14} condition on the specific 
Star Formation Rate (sSFR):
\begin{align}
-11.8 <  \log(sSFR) < -10.8 \tag*{(green valley) }
\end{align}
independently of  
the visual classification; 
in the figures we will plot the spiral, green valley and elliptical
galaxies respectively in blue, green and red
 (notice that the division into 
separated sequences 
of {\it star-forming},
{\it intermediate} and {\it passive} galaxies is still under debate, 
see e.g. \citealt{Casado15}). 
 After selecting these subsets of galaxies from SDSS, 
the final galaxy sample
that we will use
consists of $\sim 7200$ spirals, $\sim 700$
ellipticals and  $\sim 5700$ green valley galaxies for DR4, and 
$\sim 145.000$ spirals, $\sim 45.000$ ellipticals and 
$\sim 63.000$ green valley galaxies for DR7. For the sake
of clarity, in the figures we will show 
only $\sim 10\%$ of these galaxies, but we will plot the contours 
enclosing 50\% and 80\% of objects of each type.
 

\section{Galaxy properties}

In the next sections we will compare the different physical properties of the 
simulated galaxies (magnitudes/colors, concentrations, S\'ersic indices,
stellar masses, mean stellar ages/metallicities, gas metallicities, 
SFRs) with SDSS data.
For most of the galaxy properties, we will show the effects of including 
the observational
biases in the calculation mimicking SDSS (OBS) when the galaxies
are observed in the face-on projection, and the results 
derived directly from the simulations or with little post-processing 
(SIM) as commonly used in simulation studies. Furthermore,
we will show the results derived using an intermediate approach, 
where we apply simple refinements to the SIM method that are more
directly comparable to the OBS results. 
We have also derived the properties following
the OBS method but using the edge-on projections (OBS-edge).
For the different galaxy properties, we will 
provide linear best-fit formulae of the relation between the
OBS and SIM values, as well as between OBS and the other methods, 
using linear regression.

The techniques and models used to generate mock spectra from our simulated 
galaxies and to extract their physical properties have been described 
in PaperI, where we also discussed the effects 
of the SDSS small-aperture spectrograph (fibre bias)\footnote{The 
global properties are derived considering always a Field of
View (FoV) of 60x60 kpc, see PaperI.}.  As explained below, these are
based on the conversion of the simulation's outputs into mock SEDs
obtained either using a SPS model or the radiative transfer code {\sc sunrise}.
In the next sections we will focus on showing for which properties 
it is more important to apply observational techniques when simulations 
are compared to observations, and in particular whether mimicking the 
biases of SDSS makes the simulations look closer to real 
spiral/elliptical/green 
valley SDSS galaxies.


\subsection{Magnitudes, colours and stellar masses}
\label{sec:magnitudes}

In this section we compare  the position of our simulated galaxies 
in the colour-magnitude/colour-mass 
diagrams with SDSS data, showing 
also the changes due to the different techniques applied
to calculate these quantities. 
In the figures we will use the $(u-r)$ colour, as for this colour
the SDSS galaxies show a clear bimodality in the colour-magnitude 
diagram (\citealt{Strateva01, Baldry04}). We apply the following methods to 
calculate the magnitudes (see also PaperI):
\begin{itemize}
\item {\bf{\small OBS [PETRO]}\footnote{In the following, we give in 
brackets the reference labels used in PaperI when different.}}: the 
magnitudes are derived from
{\sc sunrise} face-on images (edge-on for the OBS-edge method)
with a procedure 
that mimics the  SDSS
Petrosian magnitudes calculation \citep{Blanton01,Yasuda01}, 
i.e. extracting the Petrosian Radius \citep{Petrosian76} in the 
$r$-band, and taking the flux 
inside two Petrosian Radii in all bands to calculate the magnitudes. 
The Petrosian radius $R_P$ is the radius that for 
a galaxy with luminosity profile $I(r')$ satisfies:
$$
\frac{\int_{0.8 \,R_P}^{1.25 \,R_P} dr'2\pi r' I(r')/[\pi (1.25^2 - 0.8^2)\,R_P^2]}{\int_0^{R_P} dr' 2\pi r' I(r')/[\pi R_P^2]} \equiv 0.2
$$
The flux inside two Petrosian radii recovers nearly 98\% of the light for an 
exponential profile and $\sim 80\%$ for a DeVacouleur profile \citep{Shen03}.
For our galaxies, the fraction of $r-$band flux inside two Petrosian 
radii varies from $\sim 52\%$ (E-CS$^+$) to 
$\sim 100\%$ (C-MA), and is on average $\sim 80\%$. 
\item {\bf SIM [BC03]}: the magnitudes of the simulations in the
$(u,g,r,i,z)$-bands have been calculated with the \citet[][BC03 hereafter]{Bruzual03}
SPS model, assigning each stellar
particle a spectrum according to its age, metallicity and mass.
This is one of the most
common and fast ways to calculate the spectra/magnitudes of simulated galaxies.
\item  {\bf BC03-dust [CF00]}: we add to the spectra calculated 
with BC03 the effects of dust extinction 
assuming the \citet{Charlot00} model (CF00). CF00 uses 
different extinction curves for young stellar 
populations (supposed to be born in dusty molecular clouds) and old ones 
(extincted only by dust in the ISM), without any dependence on the
inclination of the galaxy\footnote{We note that the BC03-dust method shifts 
the results of BC03 approximately by  a constant factor, as the offset between 
different galaxies can only (slightly) change due to the different 
number of young star particles.}.
The CF00 free parameters are set according to the values given in 
\citet{DaCunha08}, which are slightly different from the ones 
derived by CF00 fitting a set of star-forming galaxies;
the results of the BC03-dust model then depend somehow on the values assumed
for the free parameters, and cannot be considered fully-predictive (see 
also PaperI).

\end{itemize}

For the calculation of the total stellar masses we follow these procedures:
\begin{itemize}
\item {\bf{\small OBS [PETRO]}}: the masses are derived fitting the 
Petrosian magnitudes in all five photometric $(u,g,r,i,z)$-bands in 
the face-on view (edge-on in OBS-edge)
to the grid of models described in \citet{Walcher08}, after subtracting
 nebular emission, as the fitted models include only stellar 
light\footnote{To remove the nebular contribution from the broad-band 
magnitudes we calculate the relative contribution of nebular emission 
within the fiber in each photometric band fitting the fiber spectrum with 
the {\sc starlight} code, and 
assume that the relative contribution of nebular emission for the total
galaxy is the same as in the fiber (see PaperI for details).}.
\item {\bf {\small {SIM}}}: the total stellar mass is calculated summing
the mass of star particles (within the 60x60 kpc FoV), extracted directly from the simulations' snapshots. 
 Note that the mass in stellar particles 
inside two Petrosian radii in our simulations 
is on average $\sim 85\%$ of the total stellar mass 
in the FoV, ranging 
from $\sim 60\%$ (B-CS$^+$) to $\sim 100\%$ (C-MA).
\end{itemize}

\begin{figure}
  \centering
\includegraphics[width=7.25cm]{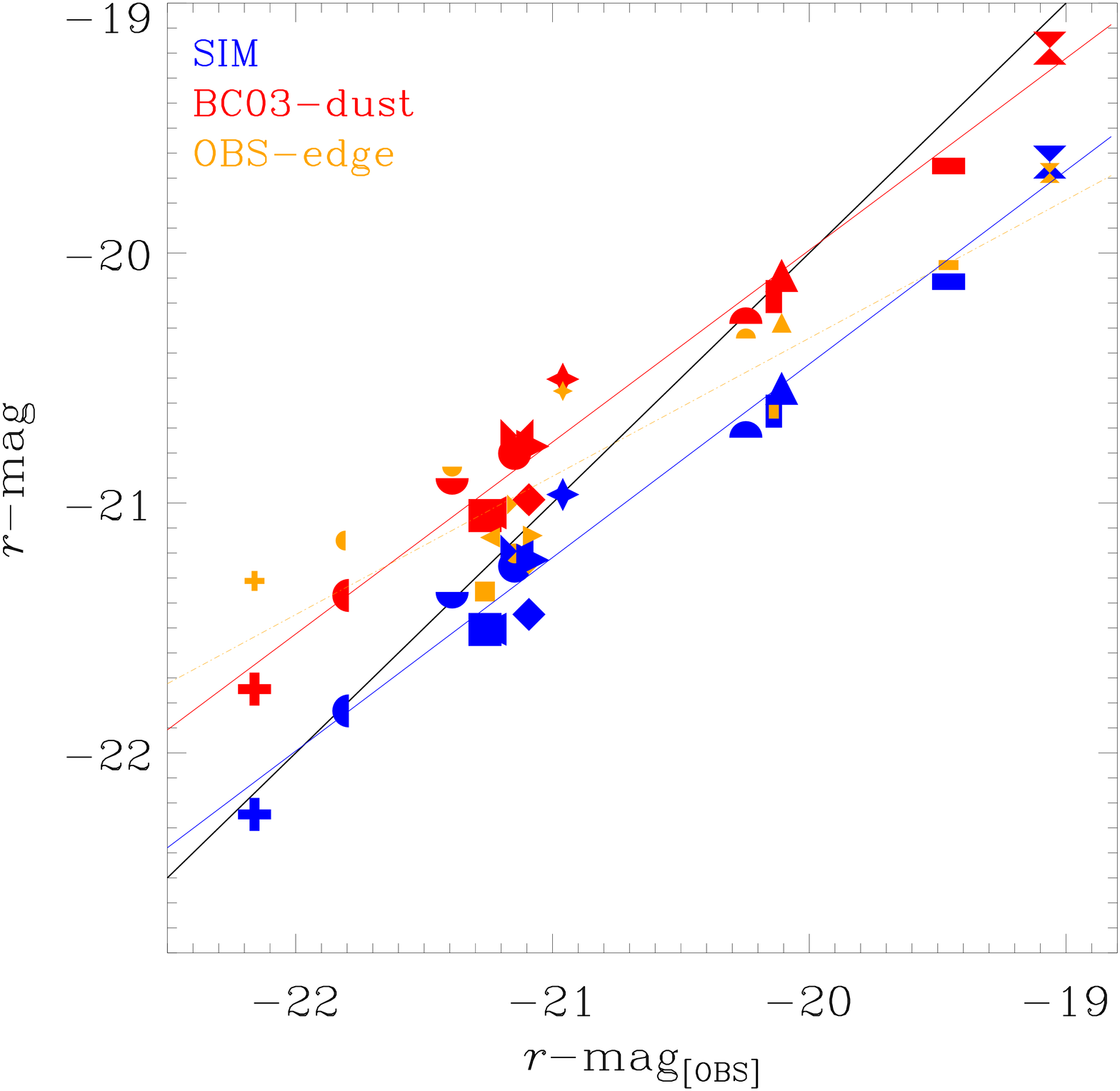}
\includegraphics[width=1.025cm]{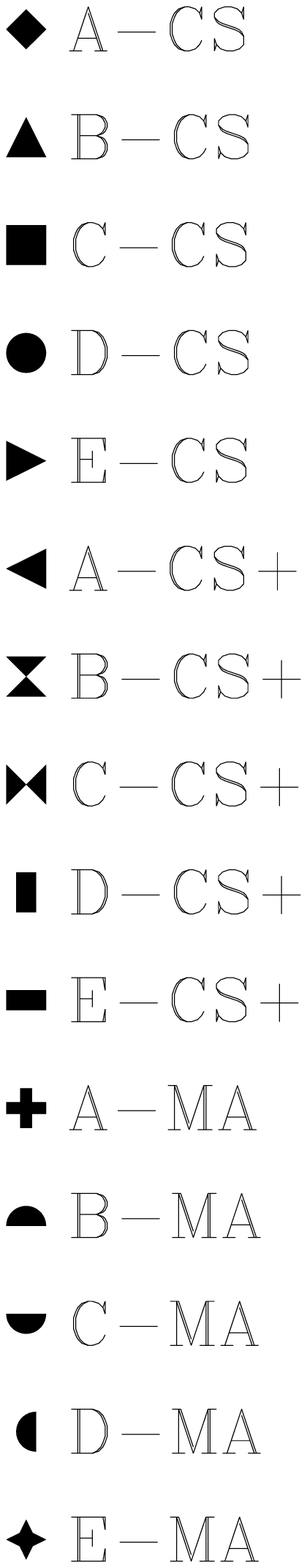}
\caption{$r$-band absolute magnitudes of the simulated galaxies using the
SIM, BC03-dust and OBS-edge methods, plotted against the observational value
OBS in the 1-to-1 relation
(black solid line). The blue, red and orange lines are, respectively, the linear
fits of the SIM, BC03-dust and OBS-edge points.}
\label{fig:1_to_1_mag}
\end{figure}

\begin{figure}
  \centering
\includegraphics[width=7.25cm]{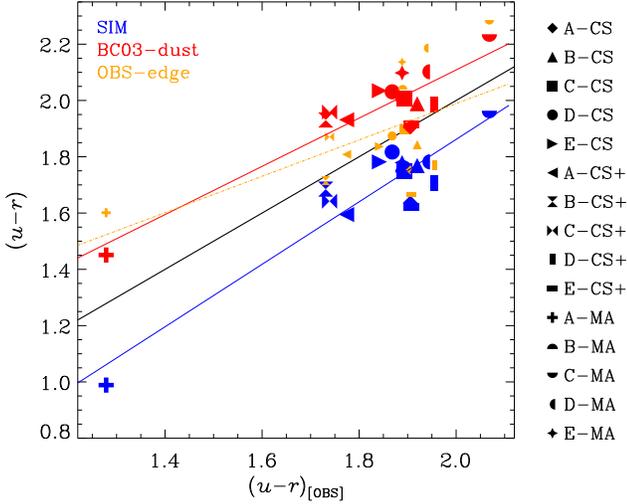}
\includegraphics[width=1.025cm]{figures/magnitudes_label_vertical.eps}
\caption{$(u-r)$ colours of the simulated galaxies using the SIM,
BC03-dust and OBS-edge methods, as a function of the corresponding 
values obtained
with OBS.
We also show the $1-$to$-1$ relation (black line) 
and linear fits of SIM, BC03-dust and OBS-edge (blue, red and orange lines 
respectively).}

\label{fig:1_to_1_col}
\end{figure}

In Fig.~\ref{fig:1_to_1_mag} we show the one-to-one relation of the  
SIM/BC03-dust/OBS-edge 
versus OBS ($r$-band) magnitudes, and the
results of linear fits to the relations. From the figure we can see that 
the SIM method gives in general lower (brighter) magnitudes compared 
to OBS, in particular for the fainter galaxies ($r$-mag $\greatsim -21$),
while for the brighter objects ($r$-mag $\lesssim -21$) it agrees 
better with the OBS estimation.
These discrepancies are the result of the composite effect of 
dust extinction 
-- dust is not considered in SIM and hence  gives brighter
magnitudes than OBS --
and cutting the luminosity profile at $2 \,R_P$
-- OBS uses the Petrosian magnitudes, further reducing the total flux
(as it misses 
the external part of the galaxy profile,
the effect being larger for galaxies with non-exponential profiles).
Note that these results strongly depend on the amount of dust, 
the orientation (face-on/edge-on), and the luminosity profile
of each galaxy. 
In fact, when the galaxies are seen edge-on (OBS-edge), the 
derived magnitudes
are different compared to those found from the face-on images.
The differences are
due to the effect of  dust which in general affects more the 
edge-on projections for dust-rich galaxies,
and to the different values of the Petrosian radii, which are much 
higher when we see the  galaxies edge-on  (in particular
 for B-CS$^+$, D-CS$^+$ and 
E-CS$^+$), resulting in most of the cases in higher (i.e. fainter) 
magnitudes compared to OBS.

The BC03-dust model gives in most of the cases galaxies with 
fainter $r$-band magnitudes compared to OBS. 
In this case, it is important to note that BC03-dust 
is based on an angle-averaged dust model, from which we expect fainter 
magnitudes compared to the face-on magnitudes from OBS (at least 
for galaxies with sufficiently large $R_P$).
 For galaxies of intermediate brightness, the OBS values lie in between 
those given by the SIM and BC03-dust methods. 
Note that, as explained above, the SIM and BC03-dust models give a similar
slope, 
indicating that the shift caused by the inclusion of dust
in the magnitudes of galaxies (in the range analysed here) is approximately
the same for all of them. 

We have performed a linear fit to the relations between the 
SIM/BC03-dust/OBS-edge and the
OBS  $r$-magnitudes, and found
the following values for the slope, zero point
and correlation 
coefficient $R$:  
\begin{gather*}
r-\text{mag}_{\text{[SIM]}} = 0.77 \times \{r-\text{mag}_{\text{[OBS]}} \} - 4.96 \;\; [\text{mag}]  \\
R_{[SIM]} = 0.986 
\end{gather*}
\begin{gather*}
r-\text{mag}_{\text{[BC03-dust]}} = 0.77 \times \{r-\text{mag}_{\text{[OBS]}} \} - 4.63  \;\; [\text{mag}] \\
R_{\text{[BC03-dust]}} = 0.986 
\end{gather*}
\begin{gather*}
r-\text{mag}_{\text{[OBS-edge]}} = 0.55 \times \{r-\text{mag}_{\text{[OBS]}} \} - 9.28 \;\; [\text{mag}] \\
R_{\text{[OBS-edge]}} = 0.911 
\end{gather*}
As can be seen in Fig.~\ref{fig:1_to_1_mag}, a linear
fit is a good approximation for these relations, as quantitatively indicated 
by the high values of $R$. Note, however, that these relations (as well as
the ones that we discuss below) somehow depend
 on the specific simulation code and sub-resolution model adopted, and
the dependence of these relations on the specific implementation of 
hydrodynamics has not been fully explored yet.

From Fig.~\ref{fig:1_to_1_col} we see a similar behaviour for the $(u-r)$ 
colours of the SIM, BC03-dust and OBS-edge methods against the OBS values.
The SIM method gives, as the effects of dust are ignored, bluer 
colours, in particular for A-MA (this is a very young and star 
forming galaxy, see
Sections~\ref{sec:stellar_ages_and_stellar_metallicities}
and \ref{sec:star_formation_rate}). 
In contrast, the BC03-dust method predicts redder colours compared to OBS, as the
reddening is estimated angle-averaged. Note that the use of the
Petrosian magnitudes can also have an impact on the estimation of
the colours, in some cases with differences
reaching $\sim 0.1-0.2$ mag (galaxies
have in general different luminosity profiles/scalelengths in 
the different photometric bands, see e.g. \citealt{Fathi10}), although 
the effects are strongly galaxy-dependent.
The OBS-edge colours are in general redder compared to OBS,
although 
in some cases they appear slightly bluer. This is
due to the combined effect of the low amount of dust extinction and the 
changes in the $(u-r$) colour  due to the use of the Petrosian magnitudes.

The linear fitting functions (with respective goodness-of-fit indicators) 
obtained 
for the SIM, BC03-dust and OBS-edge results are:
\begin{gather*}
(u-r)_{\text{[SIM]}} = 1.11 \times (u-r)_{\text{[OBS]}} - 0.36 \;\; [\text{mag}] \\
R_{\text{[SIM]}} =  0.925
\end{gather*}
\begin{gather*}
(u-r)_{\text{[BC03-dust]}} = 0.86 \times (u-r)_{\text{[OBS]}} + 0.40 \;\; [\text{mag}] \\
R_{\text{[BC03-dust]}} = 0.912 
\end{gather*}
\begin{gather*}
(u-r)_{\text{[OBS-edge]}} = 0.65 \times (u-r)_{\text{[OBS]}} + 0.70 \;\; [\text{mag}] \\
R_{\text{[OBS-edge]}} = 0.587
\end{gather*}
Note that the slope of the fit is close to one for
the SIM method, but the correlation is slightly worse compared to  that
found for the $r-$magnitudes\footnote{The A-MA $(u-r)$ colours
are far from the range
covered by the rest of the galaxy sample. 
If we ignore this galaxy
for the fits, we obtain lower correlation factors ($R\approx 0.55 - 0.65$) in
all cases, with SIM slope and zero-point of 0.68 and 0.45, for the
BC03-dust method a slope 0.63 and zero-point 0.84, and in the case of OBS-edge 
a values of 1.10 and -0.17 for the slope and zero-point respectively.}.

\begin{figure}
  \centering
\includegraphics[width=7.25cm]{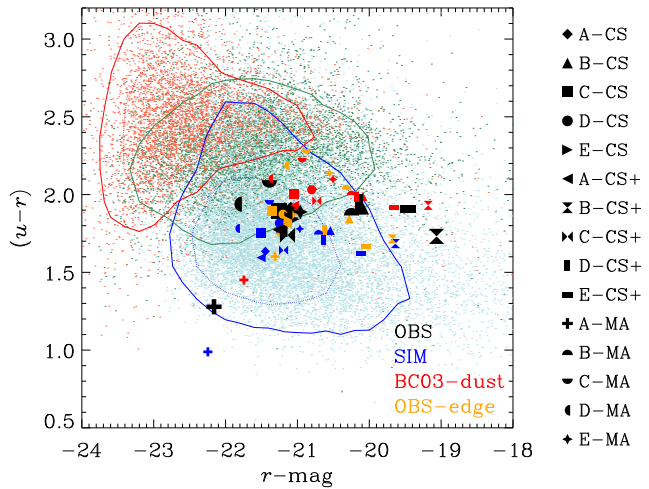}
\includegraphics[width=1.025cm]{figures/magnitudes_label_vertical.eps}
\caption{Colour-magnitude diagram of SDSS galaxies and simulations,
using different methods to calculate the magnitudes of the  
simulated galaxies. In blue, green and red are the SDSS galaxies
classified as spirals, green valley and ellipticals with
their respective contours enclosing 50\% (dotted lines) and
80\% (solid lines) of the datapoints.}
\label{fig:col_mag_SDSS}
\end{figure}

In Fig.~\ref{fig:col_mag_SDSS} we show the colour-magnitude
diagram of SDSS galaxies and the results obtained for our simulated
galaxies using the OBS method. 
From the SDSS data, we select the 
Petrosian colours/magnitudes, making it consistent with our calculation
in OBS. We also show contours for blue, red and green valley 
galaxies that enclose 50\% and 
80$\%$ of the corresponding datapoints. 
We find that  most of the simulated galaxies 
are consistent with
the photometrical properties of SDSS blue/green valley galaxies. A-MA 
is in the bluer outer part of the blue sequence, while 
E-CS$^+$ and B-CS$^+$ are outside of the region where most of the   
data are located. 
For comparison, we also show in this figure the results obtained with
the SIM, BC03-dust and OBS-edge methods.
As discussed above, using the SIM  method moves the galaxies slightly down, 
more into 
the region of the blue sequence.
Applying the BC03-dust model to calculate the magnitudes, 
the simulated galaxies move to the right (i.e. fainter magnitudes) and up 
(i.e. redder colours) towards the green valley region.
With the OBS-edge method the galaxies look in general
slightly redder and fainter,
 although for most of them the position in the diagram 
does not change significantly.

\begin{figure}
  \centering
\includegraphics[width=7.25cm]{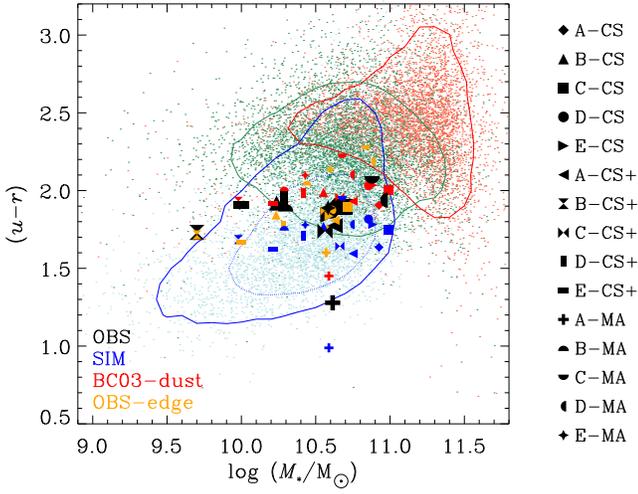}
\includegraphics[width=1.025cm]{figures/magnitudes_label_vertical.eps}
\caption{Colour-mass diagram of SDSS and simulated galaxies. The 
SIM/BC03-dust methods use the data of stellar particles of the simulation 
snapshots, while OBS and OBS-edge mimic the biases in the SDSS 
derivation of stellar mass and colours from face-on and edge-on photometric
images.
In blue, green and red are the SDSS galaxies
classified as spirals, green valley and ellipticals with
their respective 50\% (dotted) and 80\% (solid) contours.}
\label{fig:col_mass_SDSS}
\end{figure}

Fig.~\ref{fig:col_mass_SDSS} shows the colour-mass diagram,
including the SDSS datapoints and the results of the simulations
using the four methods. 
The results are similar to those found in Fig.~\ref{fig:col_mag_SDSS}, with 
most of the simulated galaxies both in the blue sequence and its intersection
with the green valley when the OBS 
method is applied, while they move slightly down (up) using the SIM 
(BC03-dust/OBS-edge) technique.
Note that the shift in stellar mass obtained applying the OBS method
is significant for 
the CS sample ($\sim 0.3$ dex), as in the CS code the mass loss of stellar 
particles due to stellar evolution is not well described (see PaperI). 
When the code includes stellar
mass loss by stars in the AGB phase (CS$^+$/MA samples),
the shift in stellar mass is less important, of the order of $\sim 0.1-0.2$
dex. Note that
the main uncertainties in the observational derivation
are related to the use of the Petrosian magnitudes and to 
the simplified procedure to construct the grid of models used in the fit,
in particular in the assumptions of star formation history (SFH) and dust 
attenuation (see e.g. \citealt{Michalowski14, Mitchell13, Wuyts09}).

In summary, we find that 
{\it including the observational biases has
limited influence on the position of the simulated galaxies
in the colour-magnitude and colour-mass
diagrams}.
As the differences between the direct results of the
simulations and those obtained mimicking the biases of SDSS data
are small, {\it the values of magnitudes and colors
derived applying SPS models to simulations can be compared
with observations at a good approximation} (at least for
old enough galaxies). 
Including a simple dust model to the direct result of simulations
does not seem to be useful to improve the comparison with observations.
The use of the different projections (face-on/edge-on) has 
small influence on the position of the galaxies in the 
colour-magnitude and colour-mass diagrams, resulting in (slightly) redder 
and fainter galaxies 
for edge-on views.
Our galaxy sample looks photometrically  
similar to galaxies in the blue sequence/green valley, and in most
of the cases is inside the range of real galaxies
in the colour-magnitude and colour-mass diagrams.

The accuracy of the stellar mass determination in observations, which
affects the positions of the galaxies in the colour-mass diagram,
has been also investigated by several previous studies.
For example, 
 \cite{Wuyts09} have shown that,
if more filters are used when fitting the photometry,
the precision in the derived stellar mass can increase
up to $\sim 0.03-0.13$ dex. 
On the other hand, stellar mass estimations based
on fitting the SED have
an accuracy of a factor of $\sim 2$ for normal star-forming galaxies 
\citep{Hayward15, Torrey15}, which can however
be improved
assuming double-component SFHs in the fitted templates \citep{Michalowski14}.
In the case of  SDSS, 
the masses derived from the SED (correcting for the limited fiber size
using the $z-$band luminosity) and from photometry have been shown
to agree within $\sim 0.2$ dex over the range 
$10^8 - 10^{12} M_{\odot}$  \citep{Drory04}.

\subsection{Concentration and S\'ersic index}

The concentration index $c$ \citep{Fraser72,Abraham94}  is defined in SDSS 
as the ratio \citep{Shen03}:
$$
c = \frac{R_{90}}{R_{50}}
$$
where $R_{90}$ and $R_{50}$ are respectively the radii including 90\% and 50\%
of the total Petrosian light.
The concentration index has been shown to correlate with the morphological 
type \citep{Shimasaku01}, and hence is a useful tool for morphological 
classification in large galaxy surveys. 

\begin{figure}
  \centering
\includegraphics[width=7.25cm]{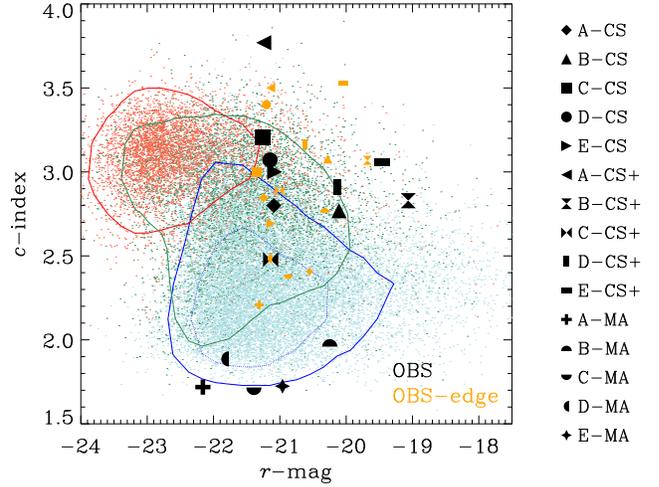}
\includegraphics[width=1.025cm]{figures/magnitudes_label_vertical.eps}
\caption{Concentration index $c$ (in the $r$-band) for SDSS 
galaxies and simulations, 
plotted versus the $r$-band Petrosian absolute magnitude, calculated
in face-on (OBS) and edge-on (OBS-edge) projections. The dotted (solid)
contours enclose 50\% (80\%) of the blue, green and 
red SDSS sample.}
\label{fig:concentration_SDSS}
\end{figure}

In Fig.~\ref{fig:concentration_SDSS} we show the concentration index $c$,
calculated in the $r$-band, as a function of
the Petrosian $r$-band absolute 
magnitude, of our simulated galaxies. 
As these
are purely observational properties, 
we only show the results 
extracted from the {\sc sunrise} face-on (OBS) and edge-on (OBS-edge) 
images. 
From this figure we see that the concentrations obtained using the face-on
images (OBS) of
the CS/CS$^+$ samples are different compared to those of the MA galaxies, the 
former being more concentrated than the latter --  with 
concentration indices larger by $\sim 0.5 - 1.5$. In fact, 
most of the CS/CS$^+$ simulations have concentrations  
consistent with SDSS
green valley galaxies; however, some them (A-CS$^+$, B-CS$^+$, 
E-CS$^+$) are outside the region  covered by the majority of 
the datapoints
(note that two of these galaxies have concentrations in the range
of observations but disagree in the magnitudes).
In particular, the  concentration index of A-CS$^+$ is larger than 3.7;
 only $\approx 0.07 \%$ of the galaxies have concentrations
above this value. 
On the other hand,
the MA galaxies have 
low concentration indices and 
lie at the bottom of the range covered by SDSS data --
only $6.5\%$ of the SDSS galaxies have concentrations below
$2$ where the 5 MA galaxies appear, and only 
$\approx 0.4\%$ below 1.8 where 3 of the MA galaxies lie.

The effect of the orientation on the estimated concentration indices 
is significant:
when we use the edge-on images, the concentrations of the three samples 
get closer, 
and most of them lie in the region of the green valley/blue sequence
galaxies, although the MA concentrations are in general lower 
than those of the CS/CS$^+$ samples. Note that the concentrations
obtained from the simulations are derived considering the 
two extreme orientations (face-on/edge-on),
while the observational sample is not corrected for inclination effects.

\begin{figure}
  \centering
\includegraphics[width=7.25cm]{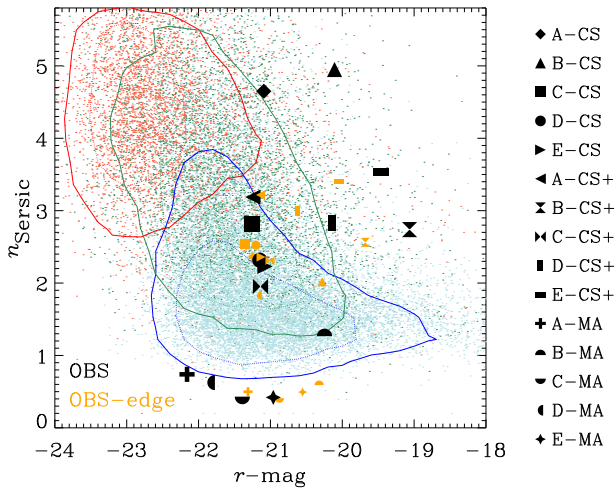}
\includegraphics[width=1.025cm]{figures/magnitudes_label_vertical.eps}
\caption{S\'ersic indices versus absolute (Petrosian) magnitudes in 
the $r$-band 
for SDSS galaxies (from \citealt{Blanton05_1}) and simulations
in face-on and edge-on views (OBS/OBS-edge respectively),
together with the 50\% (dotted) and 80\% (solid) contours.}
\label{fig:sersic_SDSS}
\end{figure}

In Fig.~\ref{fig:sersic_SDSS} we plot the $r$-band S\'ersic indices  
of the SDSS galaxies and simulations as a function of 
the Petrosian $r$-band magnitudes.
The observational data are taken from the NYU-VAGC catalogue 
\citep{Blanton05_1}\footnote{Note that the S\'ersic fit in the 
catalogue is performed on
the photometric images without considering the inclination.}, 
while the S\'ersic indices of the simulations have 
been calculated fitting
a single S\'ersic profile to the $r$-band face-on (OBS) and edge-on
(OBS-edge) images generated 
with {\sc sunrise}, using the GALFIT code 
\citep{Peng02,Peng10} and assuming arbitrary axis ratio, central pixels 
positions and angle in the fit.
In the figure we see that the samples have different 
S\'ersic indices; when the CS/CS$^+$ galaxies are observed face-on they
have in general indices $2 \lesssim n_{s} \lesssim 5$ 
and they lie in the region of spirals/green valley 
galaxies,
even though some of them are somewhat outside the area covered by
the data (B-CS, B-CS$^+$, E-CS$^+$), mainly in terms of the
$r-$magnitudes. The MA sample has face-on indices  
$n_{s} \lesssim 1$ (hence close to an exponential profile $n_{s} = 1$),
below the contour that includes 80$\%$ of the data points of
spiral galaxies. In fact, the five MA galaxies have S\'ersic index $n_s<1.4$, 
where less than 10$\%$ of the observational datapoints
are.  Furthermore, 
 four MA galaxies have $n_{s} <0.8$, which corresponds
to only $1.1\%$ of the SDSS sample. 

Similarly to what we found for the concentration
indices, we find that projection has an impact on the derivation
of the S\'ersic indices. The use of the edge-one views
causes  
the CS/CS$^+$ galaxies to have lower values compared to
those obtained from the face-on projections; as a consequence
they lie
closer to the region of the green 
valley/blue sequence galaxies. In the case of the MA sample, the 
OBS-edge method predicts higher values for $n_S$, but the galaxies are in
most of the cases still outside the 80\% contour of the SDSS spirals (for 
a discussion about the origin of these trends see e.g. 
\citealt{Maller09, Pastrav12}).

Figs.~\ref{fig:concentration_SDSS} and Fig.~\ref{fig:sersic_SDSS}
show that {\it the CS/CS$^+$ and MA samples are 
morphologically different},
with the MA galaxies lying at the low extreme of the range of spirals both 
in concentration
and S\'ersic index when these quantities are calculated face-on, 
consistent with only a small fraction of the SDSS galaxies, 
while galaxies of the CS/CS$^+$ sample lie in the 
region where spirals and green valley galaxies overlap. 
The use of the different projections has however a significant 
influence on 
the position of the galaxies both in the concentration-magnitude 
and S\'ersic index-magnitude
diagram.

\subsection{Stellar ages and stellar metallicities}
\label{sec:stellar_ages_and_stellar_metallicities}
In this section we compare the stellar ages and metallicities of the 
simulated galaxies with SDSS data.
The ages/metallicities in the SDSS-Garching DR4 are derived using 
the method described in \citet{Gallazzi05,Gallazzi06}, which is based on
simultaneously fitting different absorption features adopting a Bayesian 
inference approach. To calculate the mean ages/metallicities in our
simulations we follow these procedures:

\begin{itemize}
\item {\bf{\small OBS [LICK-IND-fibre]}}: we run {\sc sunrise} 
without nebular emission and using BC03 as input stellar model\footnote{
Since the input stellar model commonly used in {\sc sunrise} is SB99, which has
sampling $\Delta \lambda \sim 20 ~\rm{\AA{}}$ in the optical, 
the spectral resolution of the SED when the SB99 input spectra is used
is too low 
to reliably measure the Lick indices. Note also that if nebular
emission is neglected and the BC03 stellar model is used, the
interpretation of results is more direct, as the Gallazzi et al. 
method assumes BC03 in the fit, and requires the subtraction of 
nebular emission for the calculation of the indices.}; 
we select the spectra 
inside a circular region of 4 kpc radius from the center of the galaxies
both in face-on (OBS) and edge-on (OBS-edge) projections, mimicking 
the fibre size of the SDSS spectrograph at $z \sim 0.15$ (fibre FoV).
From the spectra, we measure the 
strength of the D$4000n$, H$\beta$, H$\delta_A$+H$\gamma_A$, 
[$\text{Mg}_2$Fe] and  [MgFe]' absorption features 
\citep{Worthey94,Worthey97,Balogh99} 
and we compute the mean ages and metallicities fitting these indices
with the method described in \citet{Gallazzi05}.
As the Gallazzi et al. method is also sensitive to the estimation of the 
errors on the indices, from each noiseless galaxy spectrum obtained with 
{\sc sunrise} we produce 1000 different spectra,
adding Gaussian-distributed 
random noise with S/N = 10, and we measure 1000 times the strength
of the absorption features for each galaxy in the two projections, using 
as final value for 
the indices and related errors, respectively, the average and 
standard deviation of the measurements. Note that this method to 
calculate the indices is 
different from the one presented in PaperI, as now the indices 
are more consistently extracted directly from the total stellar spectra,
and the errors on the measurements are better estimated\footnote{In fact, 
there are significant differences between our new results and those of PaperI,
that we discuss in Appendix~\ref{app:stell_age_met}.}.

\item {\bf{\small SIM [SIM-fibre]}}: the mean
ages/metallicities have been
calculated averaging the (linear) ages/metallicities of stellar particles 
in the 
fibre FoV, weighted by mass as is common in simulations studies.
\item {\bf{\small Lum-W-fibre [SIM-LUM-fibre]}}: we compute the 
mean ages and metallicites weighting, respectively, with 
the stellar particle's luminosity in the $r$-band and in all 
SDSS bands, calculated both with BC03, and considering only 
particles in the 
region sampled by the fibre.
\end{itemize}

\begin{figure}
  \centering
\includegraphics[width=7.25cm]{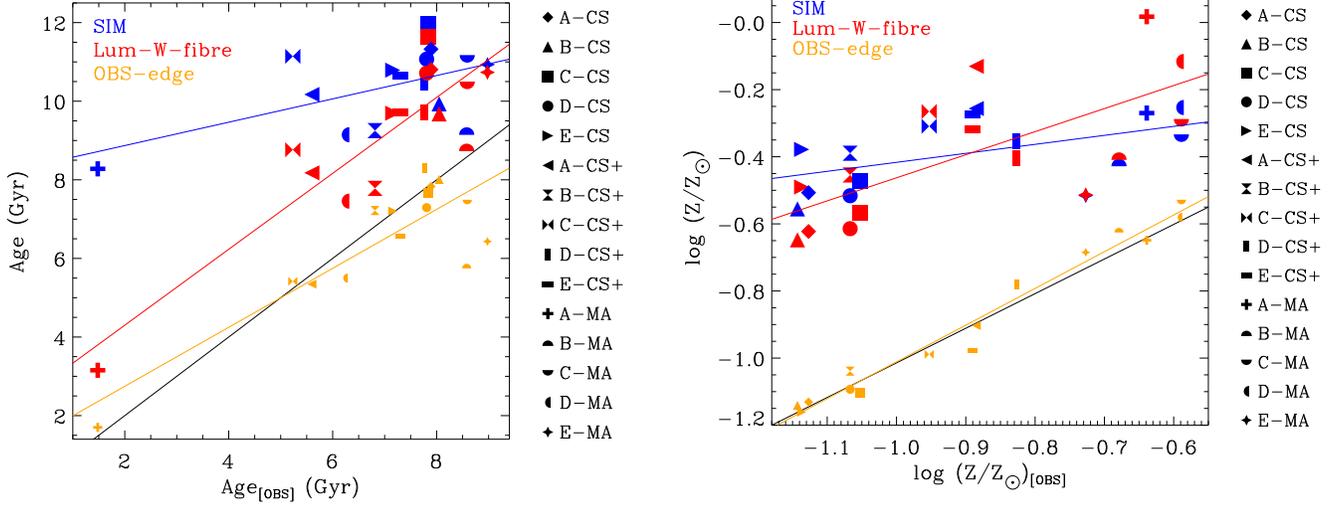}
\includegraphics[width=1.05cm]{figures/magnitudes_label_vertical.eps}
\caption{Mean stellar ages estimated with the different methods, plotted
against the observational estimations for the face-on projection, 
in the one-to-one relation (black solid line).}
\label{fig:1_to_1_age}
\end{figure}

We plot in Fig.~\ref{fig:1_to_1_age} the different estimations of 
the mean stellar ages of the simulated galaxies in the one-to-one 
relation with the value derived mimicking the SDSS observational biases,
for the face-on projection
(OBS). It is evident from the figure that the SIM values are systematically 
higher than the OBS ones, giving older stellar ages even by $\sim 2-4$ Gyr. 
When the mean age is estimated weighting with the luminosity 
(Lum-W-fibre) we obtain younger ages compared to SIM and in better
agreement with OBS, although the majority of galaxies remains older with
respect to OBS by $\sim 1-2$ Gyr. 
The discrepancy among the SIM and Lum-W-fibre methods tends to increase 
at younger
ages ($\lesssim4$ Gyr); furthermore, both relations exhibit a different slope. 
According to the OBS age estimations, the galaxies in general appear 
much younger.
In the case of the OBS-edge method,
we obtain similar values compared to OBS, although the oldest
galaxies appear slightly younger (note that the edge-on projections sample 
also part of the disks, which have in general a younger stellar content 
than the bulge).
The Lum-W-fibre method gives systematically lower ages compared to SIM,
as weighting with the luminosity gives more weight to the  
younger stellar populations, which in general emit more light than the old 
ones. 
The luminosity-weighted ages calculated with the OBS/OBS-edge methods are 
in general younger than Lum-W-fibre, due to the uncertainties 
and simplified assumptions in the procedure to construct the grid of models 
used to fit the Lick indices.
Also note that the scatter, in particular for the SIM estimation,
is relatively large, evidencing the variety of star formation histories
of the galaxies. 
It should be noted that all methods consistently 
consider the same set of stellar particles in the central part 
of the galaxies sampled by the fibre (apart from projection effects for the
OBS-edge method), so the differences are not caused by 
the presence/strength of 
age/metallicity gradients (fibre bias, see PaperI), but are purely 
related to the different techniques applied to derive the properties.

We have made  linear fits (blue/red/orange lines) to the 
SIM/Lum-W-fibre/OBS-edge datapoints obtaining 
the following relations and correlation factors:
\begin{gather*}
\text{Age}_{\text{[SIM]}} = 0.30 \times \text{Age}_{\text{[OBS]}} + 8.28 \;\; [\text{Gyr}] \\
R_{\text{[SIM]}} = 0.551
\end{gather*}
\begin{gather*}
\text{Age}_{\text{[Lum-W-fibre]}} = 0.97 \times \text{Age}_{\text{[OBS]}} + 2.37 \;\; [\text{Gyr}] \\
R_{\text{[Lum-W-fibre]}} = 0.884 
\end{gather*}
\begin{gather*}
\text{Age}_{\text{[OBS-edge]}} = 0.75 \times \text{Age}_{\text{[OBS]}} + 1.25 \;\; [\text{Gyr}] \\
R_{\text{[OBS-edge]}} = 0.850 
\end{gather*}
The scatter in the SIM datapoints is reflected 
in the low value of the 
correlation coefficient $R_{[\text{SIM}]}$.
As expected, a much larger  correlation factor is found for the Lum-W-fibre method.

\begin{figure}
  \centering
  \includegraphics[width=7.25cm]{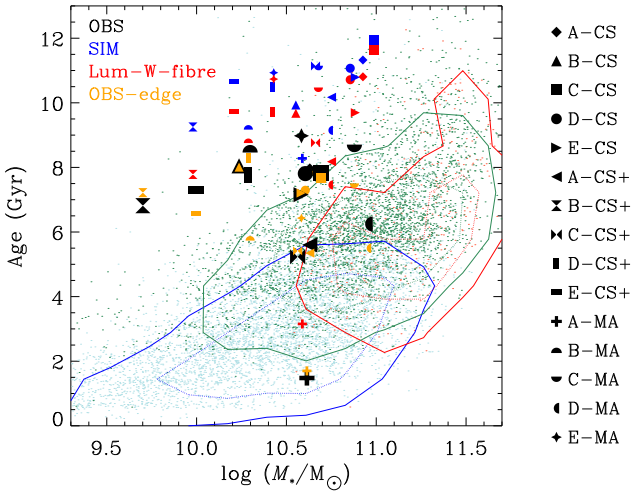}
  \includegraphics[width=1.025cm]{figures/magnitudes_label_vertical.eps}
  \caption{Stellar ages of the simulated galaxies, plotted together with 
SDSS data and the contours that enclose 50\% and 80\% of the datapoints
shown respectively as dotted and solid lines.}
\label{fig:stellar_age}
\end{figure}

In Fig.~\ref{fig:stellar_age} we compare the simulations'
stellar ages obtained with the OBS method
with mean ages from SDSS, in the stellar age-stellar
mass diagram 
(note that the stellar mass of the OBS method 
is the PETRO mass estimation, sec.~\ref{sec:magnitudes}).
Our results  show that about half of the 
simulated galaxies look older compared to the 
observations, and the rest is close or inside the contours
corresponding to green valley galaxies (CS sample) or
well inside the blue sequence (A-CS$^+$, C-CS$^+$, A-MA, D-MA) where it
intersects with green valley and red galaxies. When the OBS method
is applied to edge-on spectra in the fibre (OBS-edge), we obtain
similar results compared to OBS, while some simulations move into the
region covered by the observational  data (B-MA, C-MA, E-MA).
For reference, we also include in the figure results for the
SIM and Lum-W-fibre methods; note that in these cases 
the stellar mass
is calculated as the sum of the mass of stellar particles of the
simulated galaxies.
As discussed above, using the SIM method
the galaxies appear too old compared to the observations,
while weighting 
with the luminosity (Lum-W-fibre) moves the 
points down towards the range of observations, but
still most of the galaxies are too old compared to the real ones.

\begin{figure}
  \centering
 \includegraphics[width=7.25cm]{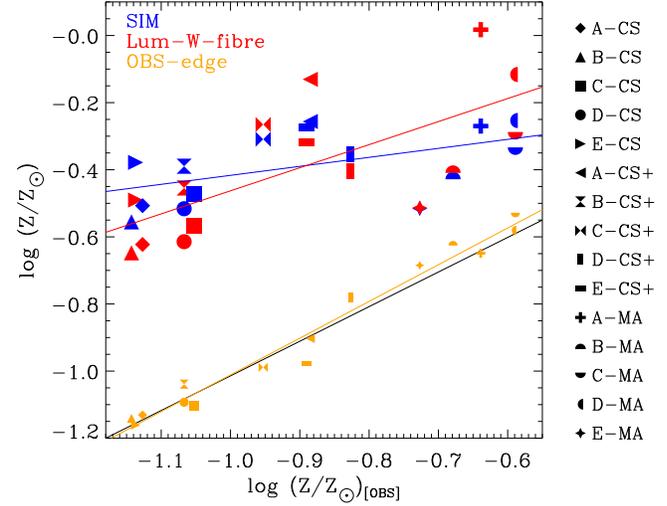}
  \includegraphics[width=1.025cm]{figures/magnitudes_label_vertical.eps}
\caption{One-to-one relation of the different mean stellar metallicity
estimations (together with the best-fit models in blue, red and orange lines) versus the method 
closest to SDSS applied to face-on spectra (black line).}
\label{fig:1_to_1_met}
\end{figure}

We make a similar analysis for the mean stellar metallicities, showing
in Fig.~\ref{fig:1_to_1_met} the comparison among the different methods.
From the figure we see that both the SIM and Lum-W-fibre methods 
give systematically 
higher metallicities compared to OBS, with the offsets 
increasing (up to $\sim 0.4-0.5$ dex) for metal poor galaxies. 
The discrepancy between SIM and Lum-W-fibre is explained
by the different weight of old and young stars when the average metallicity
is calculated weighting with the luminosity, while OBS (OBS-edge) results 
(for which 
the mean metallicity is computed luminosity-weighted) suffer from the 
uncertainties intrinsic in the method.
When the observational method is applied to edge-on spectra we obtain
very similar results compared to OBS over the full range of metallicities.

The values obtained for the linear fit and correlation coefficient are:
\begin{gather*}
\log (Z/Z_{\odot})_{\text{[SIM]}} = 0.27 \times \log (Z/Z_{\odot})_{\text{[OBS]}} - 0.15\;\; [\text{dex}]  \\
R_{\text{[SIM]}} = 0.525 
\end{gather*}
\begin{gather*}
\log (Z/Z_{\odot})_{\text{[Lum-W-fibre]}} = 0.69 \times \log (Z/Z_{\odot})_{\text{[OBS]}} + 0.23 \;\; [\text{dex}]  \\
R_{\text{[Lum-W-fibre]}} = 0.701  
\end{gather*}
\begin{gather*}
\log (Z/Z_{\odot})_{\text{[OBS-edge]}} = 1.1 \times \log (Z/Z_{\odot})_{\text{[OBS]}} + 0.08 \;\; [\text{dex}] \\
R_{\text{[OBS-edge]}} = 0.987  
\end{gather*}
The correlations are similar to the ones found for the stellar ages, with the
Lum-W-fibre method increasing the value of $R$, and the OBS-edge method
having $R\sim 1$.

\begin{figure}
  \centering
  \includegraphics[width=7.25cm]{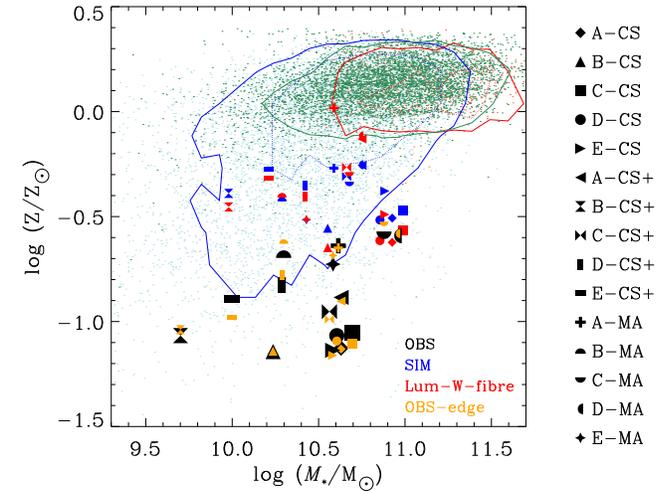}
  \includegraphics[width=1.025cm]{figures/magnitudes_label_vertical.eps}
  \caption{Mean stellar metallicity of SDSS galaxies and 
simulations, estimated using different techniques. The simulated galaxies
appear in the area of young spirals (the blue contours contain 
50\% and 80\% of SDSS spirals), but some of them are
outside the range of real galaxies.}
\label{fig:stellar_met}
\end{figure}

In  Fig.~\ref{fig:stellar_met} we show a comparison of the stellar metallicities
obtained with the OBS method and observational results.
We find that
some of the galaxies are in the area of metal-poor spirals, 
while most lie outside of the region where most of the observations
are (in particular the CS sample), with log$(Z/Z_\odot) < -0.9$.
Note that only
$\approx 1.6\%$ of the SDSS galaxies have metallicities 
lower than this value.
In the case of stellar metallicities derived using the SIM or Lum-W-fibre 
methods, galaxies appear slightly more metal-rich, with 
most of them lying
in the blue sequence, although the CS sample is again
outside the region covered by SDSS galaxies.

In conclusion, {\it we find that
the effects of using different methods to calculate 
the mean stellar ages/metallicities are strong and affect significantly
the comparison of simulations with observations}. When simple
derivations of stellar ages and metallicities done in simulation
studies are used,  the galaxies 
appear  older and more metal-rich compared to 
results obtained following observational techniques.
In the case of stellar ages, the discrepancies between SIM and OBS values 
are large, in particular at younger ages.
Weighting with the luminosity to obtain the
stellar ages also
affects the results, which are closer to the observational values but
shifted compared to OBS 
by almost a constant factor.
In the case of stellar metallicities, the discrepancy between SIM and
OBS increases at lower metallicity, while the offset is almost constant when 
the mean metallicity is calculated weighting with the luminosity.
The effect of the projection, estimated applying the observational method
to edge-on spectra, is secondary compared to the differences arising
from the use of different derivation methods.
Our results show that the majority of our
simulated galaxies appear older than real spirals,
and with metallicities similar or lower than the most metal-poor 
spirals in SDSS.

It should be noted that comparing the Lum-W-fibre and 
OBS results (both providing the luminosity-weighted ages/metallicities)
the observational method shows a bias to systematic younger 
ages and lower metallicities. Notice also that our Lum-W-fibre results
have a scattering with respect to the Lum-W-fibre/OBS relation of 
$\sim 0.3$ dex in ages and
$0.2 $ dex in metallicities, similar to the errors of the method  
estimated by \citet{Gallazzi05} in case of good signal-to-noise (i.e. 
S/N$ > 20$), namely $\sim 0.2$ dex for the ages and $\sim 0.3$ dex for the 
metallicities.
In PaperI we have shown that SED fitting methods 
are able to reach a higher accuracy in stellar age and metallicity 
determination, constraining the ages by 
$\sim 0.06$ dex and the metallicities by $\sim 0.15 - 0.25$ dex, although
still with some trends (see PaperI);
other SED fitting studies (e.g. \citealt{Cid_Fernandes05}) claim results
similar to our findings (in the case of good S/N), with ages constrained 
by $\sim 0.08$ dex and metallicities by $\sim 0.1$ dex.

\subsection{Gas metallicities}

In the SDSS-Garching DR7 dataset, the gas oxygen abundances (metallicities) 
are extracted using the method described in \citet[][T04]{Tremonti04} (see 
also \citealt{Brinchmann04}), which is based on the simultaneous fit of
different emission lines according to the \citet{Charlot01} model (CL01).
The authors also give a calibration of the $R_{23}$-metallicity 
relation -- so-called T04 calibration -- which is valid on the upper branch.
In this work we derive the gas metallicities of the simulated galaxies
in the following ways:

\begin{itemize} 
\item {\bf{\small OBS [T04-fibre]}}: the gas metallicity 
is calculated applying the T04 calibration to {\sc sunrise} 
face-on spectra (edge-on for OBS-edge)
extracted inside a circular region of 
4 kpc radius at the center of the galaxy (fibre FoV, 
sec.~\ref{sec:stellar_ages_and_stellar_metallicities}), 
after correcting for 
dust extinction with the Calzetti law \citep{Calzetti94}.
Since the T04 calibration is only valid in the
upper branch of the relation, from our sample of fifteen 
galaxies we are able to include twelve objects (eleven for OBS-edge), 
selected according to the [NII]/[OII] ratio \citep{Kewley08}.
\item {\bf{\small SIM [Mass-W]}}: we calculate the mean 
$12 + \log (O/H)$ abundance from the oxygen/hydrogen ratio of 
each gas particles, weighted by the particle's mass.
\item {\bf{\small Sim-fibre}}: 
the same as SIM, but only considering gas particles inside 
the fibre FoV in the  face-on orientation.
\end{itemize}

\begin{figure}
  \centering
\includegraphics[width=7.25cm]{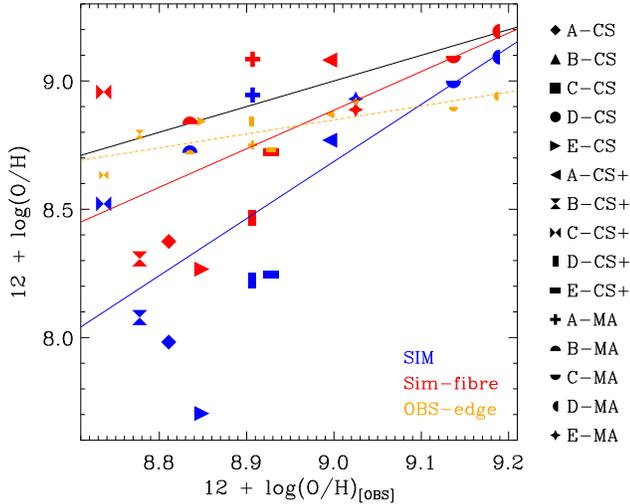}
\includegraphics[width=1.025cm]{figures/magnitudes_label_vertical.eps}
\caption{Comparison of the gas metallicities extracted using the
different methods, in the one-to-one relation (black solid line)
with the metallicity derived
applying the T04 calibration to the emission lines ratios of the simulated
{\sc sunrise} face-on (OBS) and edge-on (OBS-edge) spectra.}
\label{fig:1_to_1_gas_met}
\end{figure}

The results of these different techniques are shown in 
Fig.~\ref{fig:1_to_1_gas_met}, plotted in the one-to-one 
relation with the OBS method (T04-fibre).
The plot reveals a large scatter among the methods, particularly
for  galaxies  with $12 + \log (O/H)\lesssim 9$.
Deriving the metallicities with the SIM method gives in most of 
the cases 
lower values compared to OBS/Sim-fibre/OBS-edge
(note that the OBS/Sim-fibre/OBS-edge methods measure the metallicity in 
the metal-enriched central part of the galaxies, and that our CS and CS$^+$ 
samples have stronger 
metallicity gradients compared to MA, see PaperI).

Using the Sim-fibre method moves the metallicities closer to
OBS, even though with some scatter and with the tendency to underestimate
the metallicity of metal-poor galaxies. 
Although the Sim-fibre and OBS methods sample the same central
region of a galaxy, a discrepancy among the two methods 
is somehow expected as the emission line ratio from which 
the OBS values are extracted are based on the {\sc mappings III} code,
and the several uncertainties and assumptions on 
modelling nebular emission in the photoionization code may affect 
the derivation of the gas metallicities (see \citealt{Groves04}).

The OBS-edge method gives similar results compared to OBS;
however,
the gas metallicity of the most metal-rich galaxies is systematically
lower by $0.1-0.2$ dex compared to the results when the galaxies are
observed face-on (OBS). 
These differences may indicate that the use of a different orientation will 
affect the region sampled by the fiber due to projection effects, which on its
turn will affect the metallicity estimation. Note also that 
the uncertainties in the dust corrections for edge-on/face-on 
galaxies may also influence the determination of the gas 
metallicities.

We fitted the relations between the SIM/Sim-fibre/OBS-edge and the OBS
methods with linear functions (blue, red and orange lines, respectively),
and obtained the following parameters:
\begin{gather*}
12 + \log (O/H)_{\text{[SIM]}}=  2.22 \times \{12 + \log (O/H)_{\text{[OBS]}}\} 
- 11.33 \\
R_{\text{[SIM]}} = 0.674 
\end{gather*}
\begin{gather*}
12 + \log (O/H)_{\text{[Sim-fibre]}}=  1.50 \times \{12 + \log (O/H)_{\text{[OBS]}}\} - 4.65 \\
R_{\text{[Sim-fibre]}} = 0.621 
\end{gather*}
\begin{gather*}
12 + \log (O/H)_{\text{[OBS-edge]}}=  0.54 \times \{12 + \log (O/H)_{\text{[OBS]}}\} + 3.97 \\
R_{\text{[OBS-edge]}} = 0.804 
\end{gather*}
Note the different slopes of the relations, particularly in
the case of the SIM and
Sim-fibre methods, which also have a similarly low  correlation factor.
As expected, the  relation between the OBS-edge and OBS datapoints
has a higher correlation factor. 

\begin{figure}
  \centering
\includegraphics[width=7.25cm]{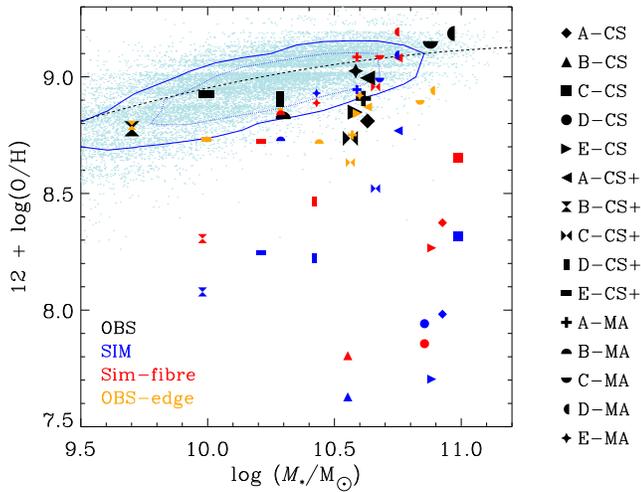}
\includegraphics[width=1.025cm]{figures/magnitudes_label_vertical.eps}
\caption{Mass-metallicity diagram for SDSS galaxies and simulations. 
The dashed line is the analytic mass-metallicity relation from 
\citet{Tremonti04}, while the blue contours encloses respectively 50\% 
(dotted) and 80\% (solid) of the datapoints. 
OBS and OBS-edge methods closely mimic SDSS biases, 
using the T04 calibration applied to the spectra inside the fibre FoV
respectively in face-on and edge-on projections.}
\label{fig:SDSS_gas_met}
\end{figure}

In Fig.~\ref{fig:SDSS_gas_met} we compare  the
gas metallicities of the simulated galaxies obtained
with the OBS method and the SDSS dataset.
We find for face-on views that most of the sample is in good 
agreement with the observations and inside the area covered by the data,
even though the majority of the galaxies have metallicities slightly 
below the T04 analytical relation (dashed line), and the
A/E-CS and C-CS$^+$ galaxies are outside the contour containing 
80\% of the data. 
For metallicities derived from edge-on spectra, all the galaxies are below
the analytic relation, and only four galaxies are inside the
80\% contour.
Note that, in the gas metallicity-stellar mass plane, 
following observational techniques makes the
galaxies more consistent with observations,
by shifting them both
in the metallicity and stellar mass values, compared to the
common estimations done in simulation studies.
For reference we also show results for the SIM and Sim-fiber; 
for SIM and Sim-fiber galaxies appear too metal-poor 
compared to observational
data. The discrepancies are large, in some cases
even by more than 1 dex.
Note also that when we take into account only particles inside the fibre FoV 
(Sim-fibre), the metallicities of 
the CS/CS$^+$ samples significantly increase 
(while those of the MA galaxies remain similar, since these galaxies 
have flatter 
metallicity gradients as shown in PaperI).

Our results show that, {\it to compare gas metallicities of simulated and real
galaxies it is important to apply to the simulations
the same
methods and calibrations than in observations, in order to make
the comparisons reliable}. An intermediate step of
obtaining a more comparable but simple gas metallicity estimation
from the simulations is to mimic the most relevant
biases of the survey, in particular the  SDSS fibre size.
The use of face-on or edge-on views has also an influence on the 
gas metallicity estimation, as the fiber may sample different regions
in a galaxy due to projection effects.
We have shown that {\it simple calculations obtained directly from 
the simulations,
that neglect all the 
observational biases, can not be properly compared to SDSS observations}.
Mimicking the SDSS derivation,  
our galaxies are close to the gas metallicities of spirals in SDSS, 
although with the trend of having oxygen abundances slightly lower than 
observational results.
This is related to the particular calibration adopted (T04), as different 
studies 
(e.g. \citealt{Kewley08}) have shown that the effects of the metallicity 
calibration used to determine the oxygen abundance are strong, affecting 
the determination with offsets up to $\sim 0.7$ dex comparing theoretical (such 
as T04) and empirical calibrations, with the theoretical 
calibrations giving in general higher metallicities (see also PaperI)

\subsection{Star formation rates}
\label{sec:star_formation_rate}

We analyse in this section the SFRs of our
simulated galaxies, and compare them to the  SDSS-Garching data.
The method used to calculate the total SFRs from the SDSS spectra is described 
in \citet{Brinchmann04} and is based on the CL01 model, 
correcting for the limited fibre size of the spectrograph 
with the technique described in \citet{Salim07}. 
For our simulated galaxies, we estimate the SFRs following these procedures:
\begin{itemize}
\item {\bf{OBS [H$\alpha$]}}: we extract the H$\alpha$-luminosity
L(H$\alpha$) from 
the  {\sc sunrise} face-on (edge-on for the OBS-edge method) spectra that, 
after correcting 
for dust
extinction with the Calzetti law using the H$\alpha$/H$\beta$ ratio, 
we convert into SFR according to the Kennicutt calibration, taking into
account with the factor $f_{\rm IMF}=1.5$ the use of Kroupa/Chabrier IMF 
\citep{Calzetti09}:
\begin{equation*}
{\rm SFR}\;({\rm M}_\odot \, / \,{\rm yr}) = 7.9 / f_{\rm IMF}  \times 10^{-42} \, L(H\alpha) \, ({\rm erg \, / \,s}) 
\end{equation*} 
Note that this method, although different from the one used in SDSS 
analysis, has been shown in \citet{Brinchmann04}
 to be in good
agreement with it, at least for our range of stellar masses. 
Note also that, in addition,  
the method is sensitive
(together with the BC03-ionizing flux) to the emission from 
young massive stars with lifetimes $\lesssim 10$ Myr \citep{Calzetti08}, 
while the SFR derived with the SIM method is averaged over a larger timescale (0.2 Gyr).
\item {\bf{\small SIM}}: we calculate the SFR directly from the simulation's
snapshots, considering the amount of total stellar mass formed over 
a certain time interval, that we set to the last 0.2 Gyr.
\item {\bf{\small BC03}}: we convert the rate of ionizing 
photons $Q(H^0)$ calculated with BC03 into SFR according to the 
calibrations given in \citet{Kennicutt98}:
\begin{equation*}
{\rm SFR}\;({\rm M}_\odot \, / \,{\rm yr}) = 1.08 / f_{\rm IMF} \times 10^{-53} \, Q(H^0) \, ({\rm s}^{-1}) 
\end{equation*} 
\end{itemize}

\begin{figure}
  \centering
\includegraphics[width=7.25cm]{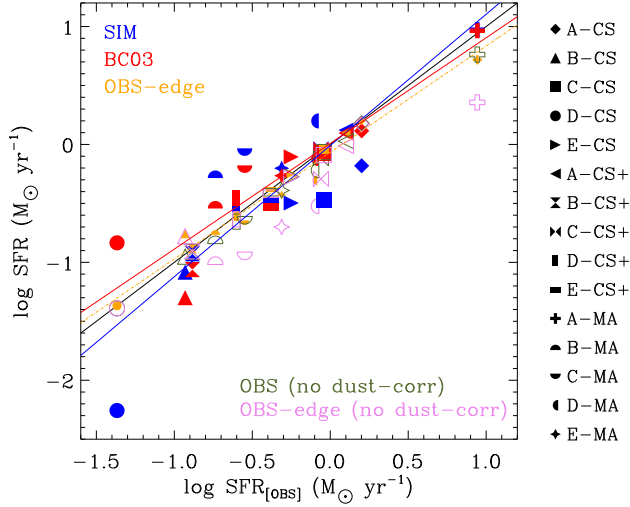}
\includegraphics[width=1.025cm]{figures/magnitudes_label_vertical.eps}
\caption{Comparison of the different SFR estimators, in the one-to-one 
relation (black line) with the observational method (OBS) based 
on H$\alpha$ luminosity. The best-fits of both SIM and 
BC03-ionizing flux (solid blue and red lines) are in good 
agreement with the relation, as well as the values derived
applying the observational method in the edge-on projection (OBS-edge). 
The results not corrected for dust show the effect of dust 
extinction on the spectra for these simulations.}
\label{fig:1_to_1_sfr}
\end{figure}

In Fig.~\ref{fig:1_to_1_sfr} we show the estimations of the
SFR using the different methods in the one-to-one relation with the OBS 
results. We find in general a tight agreement among 
them, with scatter of the order of $\lesssim 0.2-0.4$ dex, 
and only the lowest-SFR galaxy 
(D-CS) has a significantly different SIM value compared to 
the OBS estimator. 
Projection effects do not strongly affect the derived SFRs, as 
evidenced by the similar relation found in the case of the OBS-edge method 
(note that both OBS and OBS-edge are corrected for dust extinction, and 
sample the full field of view of $60 \times 60$ kpc).

The linear functions that best-fit the SIM/BC03/OBS-edge datapoints, and
the values of the correlation factors $R$, are:
\begin{gather*}
\text{SFR}_{\text{[SIM]}} = 1.11 \times \text{SFR}_{\text{[OBS]}} - 0.01 \;\;\; [\text{M}_\odot / \text{yr}] \\
R_{\text{{SIM}}} = 0.871
\end{gather*}
\begin{gather*}
\text{SFR}_{\text{[BC03]}} = 0.90 \times \text{SFR}_{\text{[OBS]}} + 0.01 \;\;\; [\text{M}_\odot / \text{yr}] \\
R_{\text{{BC03}}} = 0.921 
\end{gather*}
\begin{gather*}
\text{SFR}_{\text{[OBS-edge]}} = 0.90 \times \text{SFR}_{\text{[OBS]}} - 0.07 \;\;\; [\text{M}_\odot / \text{yr}] \\
R_{\text{{OBS-edge}}} = 0.987 
\end{gather*}
Note that the BC03 points appear to be in a better agreement with
OBS compared to SIM, with slightly smaller scatter (i.e. higher correlation 
coefficient).

It is worth noting that both BC03 and OBS are based on the conversion of 
the rate of ionizing photons into SFR and hence sample the same timescale
of star formation, 
while for the SIM method we consider a much longer timescale, 
as star formation is treated stochastically in the simulations.
This however may result in 
significant discrepancies with the observational estimators, in 
particular in the presence of recent starbursts
\citep{Sparre15}. We have tested this effect using
a timescale of 10 Myr for the SIM method; in this
case we obtain a similar fit, with slope and zero points of
$1.103$ and $- 0.061$, respectively, 
but a higher correlation factor of  $R_{\text{{SIM}}}= 0.973$.

In Fig.~\ref{fig:1_to_1_sfr}, we additionally show
the SFRs derived from H$\alpha$ without correcting 
for dust, to provide a visual impression of the amount of dust extinction
in our simulations, both in the face-on and edge-on projections.
We see that that for most of our galaxies the effects of dust are small
on the face-on spectra, while dust affects the edge-on views more significantly,
particularly for the galaxies with higher metal content (MA sample).

\begin{figure}
  \centering
\includegraphics[width=7.25cm]{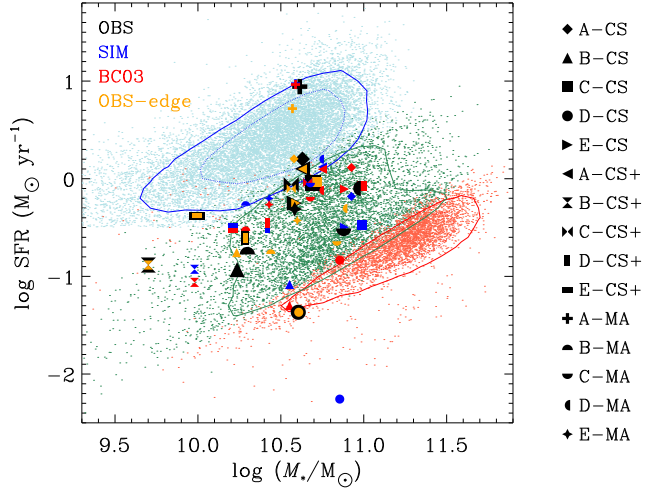}
\includegraphics[width=1.025cm]{figures/magnitudes_label_vertical.eps}
\caption{SFR-stellar mass diagram for SDSS and simulated galaxies, where the 
SDSS galaxies are shown in blue, green and red according to morphological
classification, with their respective 50\% (dotted) and 80\% (solid)
contours. For each 
simulation we plot the results of the different
methods.}
\label{fig:sfr_SDSS}
\end{figure}

In Fig.~\ref{fig:sfr_SDSS} we compare the SFRs of simulated
galaxies obtained with the OBS method to observations.
The majority of the simulated galaxies have SFRs consistent with those of
green valley galaxies,
and  only A-MA and E-CS$^+$ are in the area of the actively star-forming 
spirals. On the contrary, B-CS and D-CS have low SFRs, closer to the ones 
of red ellipticals. 
As shown in the previous figure, the position
of the galaxies in the SFR-stellar mass diagram when different
methods are applied are similar, 
although in some cases the use of the OBS method 
(which includes the Petrosian masses) 
moves the simulations towards the range of real galaxies (D-CS), or 
from the red sequence to the green valley (e.g. B-CS) due to
the different timescales over which the SFR is derived with the OBS and SIM
methods. 

\begin{figure}
  \centering
\includegraphics[width=7.25cm]{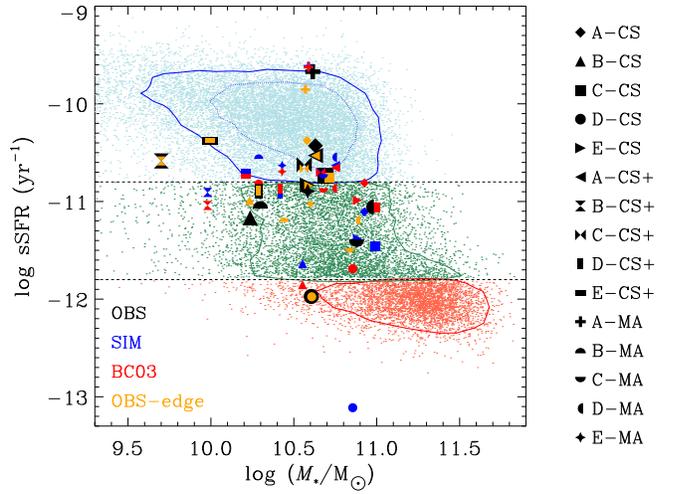}
\includegraphics[width=1.025cm]{figures/magnitudes_label_vertical.eps}
\caption{Specific SFR$-$stellar mass diagram showing SDSS data and the
50\% and 
80\% contours of each morphological type (spirals, green valley galaxies
and ellipticals
respectively in blue, green and red),
together with the values extracted from the simulated galaxies using  
different techniques.}
\label{fig:ssfr_SDSS}
\end{figure}

The Specific Star Formation Rate (sSFR) diagram 
(Fig.~\ref{fig:ssfr_SDSS}) confirms the trends of Fig.~\ref{fig:sfr_SDSS}, 
with most of the galaxies in the transition between the green valley and
the blue sequence, although now more galaxies appear in 
the blue sequence area
(note that 
the diagram is divided in three regions by the definition of green valley 
galaxies, see 
Sec.~\ref{sec:observations}).

We conclude that {\it the estimation of the SFR in simulations 
is not strongly affected by which method is used, and only mildly by the projection
if the spectra are corrected for dust extinction}; 
the values extracted
directly from the simulations can be meaningfully compared with 
observations, at least for normal star-forming galaxies.
Most of our simulated galaxies have SFRs at the 
transition between the green valley/blue sequence of SDSS galaxies. 
The good agreement between the star formation rate in simulations and
the one extracted from an observational indicator such as the H$\alpha$ 
flux is in general found for other star formation rate proxies, such as the 
[OII] line intensity \citep{Jonsson10} or the IR luminosity 
\citep{Hayward14, Hayward15}, 
at least for quiescent star-forming galaxies.

\section{Discussion and conclusions}

We have made an unbiased comparison between simulated and observed galaxies,
converting the simulation outputs into synthetic observations and applying
observational techniques to derive the galaxies' magnitudes, colours,
stellar masses, mean stellar ages, stellar and gas metallicities and star
formation rates. 
We have used 15 hydrodynamical simulations of galaxies formed in 
a cosmological context adopting three different models
for chemical enrichment and feedback, and we compared their properties
with data from the Sloan Digital Sky Survey 
(SDSS). 
In order to extract the physical properties of the simulated
galaxies, we first created synthetic spectra,
using both a SPS model (BC03) and a full radiative-transfer code 
({\sc sunrise}) to post-process the snapshots.

In a first paper of this series \citep{Guidi15},
we compared the properties of the galaxies obtained
applying different observational methods,
and we have shown that large variations can appear, 
most notably in the case of the galaxies' ages and metallicities.
In this paper, 
we focused on the methods that mimic the SDSS techniques, in order
to make an unbiased comparison with the observational dataset.
In particular, we studied which physical properties are more 
affected by the observational biases that, in the case
of SDSS, are mainly originated by the limited fiber size, 
by the methods applied to recover ages and metallicities, 
and by the use of the Petrosian quantities that affect both the magnitudes
and stellar mass estimation. 
 
We have given simple scalings to convert the direct results of
simulations into values that can be compared with the SDSS dataset
in a reliably manner,
although in some cases (most notably in the mean stellar ages of galaxies)
the correlation has a large scatter. 
Moreover, the scalings we found might depend somehow on the particular
hydrodynamical code we used to simulate the galaxies, 
which we partially tested 
by applying three different versions of 
chemical enrichment and feedback.
In addition, our results are also sensitive
to some choices in the derivation of the observables 
(e.g. inclination, as shown by the results derived in the 
edge-on projections), and 
these caveats should be taken into account when
the provided
scaling relations are used.

We found that the
biases that appear when observational techniques
are applied 
affect differently the various galaxy properties that we studied here.
In particular, for the colours and magnitudes,
mimicking observational techniques has a small effect,
and the direct results of simulations can be reliably
compared with SDSS data.  Stellar masses derived fitting the photometry
show some  discrepancies with respect to the stellar masses in simulations,
although they are small except in the cases where 
mass loss due to stellar evolution is not 
properly modelled in the hydrodynamical codes.
In contrast,
in the case of stellar ages and stellar and gas metallicities  
the effects are stronger.
For stellar ages and metallicities,
the values of the simulations following 
observational techniques predict younger and more metal-poor 
galaxies compared to their mass-weighted values.
The discrepancy 
among the methods increases both for young and for metal-poor galaxies.
Refining the direct calculation by weighting with the luminosity 
of the stellar particles improves the mean age and metallicity determination, 
although strong trends with respect to 
the observational method appear.

For the mean oxygen
abundance of the gas, we find that
applying the SDSS metallicity calibration (T04) to the spectra
in the fiber
makes the simulated galaxies to be in much better 
agreement with observations compared to a direct calculation.
Ignoring the fibre bias makes the 
galaxies appear more metal-poor,
the effects being stronger for galaxies with steeper metallicity gradients.
This has been already investigated in several studies 
(e.g. \citealt{Kewley08, Tremonti04}), 
which found that the limited fiber size strongly affects the determination
of the gas metallicity and the shape of the mass-metallicity relation, 
particularly for galaxies with masses $M_* > 10^{10} M_{\odot}$.
Our findings  show that, when simulated spectra are not available, the 
comparison of simulations' gas metallicity with SDSS data 
can be improved mimicking the fibre size of the SDSS spectrograph. 

The determination of the star formation rates is the quantity less 
affected by the method used, and the values extracted from the simulations 
and the
H$\alpha$ flux can be meaningfully compared (if dust extinction is correctly 
taken into account).
We also found that the effects of the projection are significant for
quantities such as the concentration and S\'ersic indices, 
while projection has smaller effects when physical quantities are corrected for 
dust extinction such as in the case of gas metallicity and SFR.

Our results show that,  when an unbiased comparison with the SDSS data
is performed, our simulated galaxies:

\vspace{0.2cm}

\noindent ($i$) look photometrically similar to SDSS blue/green valley galaxies,

\noindent ($ii$)  have concentrations and S\'ersic indices 
mostly in the range of SDSS galaxies, even though the 
 different feedback codes give different results and in some cases outside the observed
range, 

\noindent ($iii$)  are in  good agreement with SDSS  ages, although most of them 
appear  older 
compared to SDSS spirals,

\noindent ($iv$)  have stellar  metallicities  
consistent with  metal-poor spirals,

\noindent ($v$) show good  agreement with observations of the gas oxygen abundances, 
even if they remain slightly more metal-poor,

\noindent ($vi$) have H$\alpha$-based SFRs in the region between the SDSS green 
valley galaxies and the blue sequence, although there are objects with 
H$\alpha$-based SFRs
both in the region of strongly star-forming spirals and
in the red sequence of passive ellipticals.

In summary, we have shown that {\it a reliable comparison between
observations and simulations requires in general the conversion
of the direct results of simulations into observationally-derived quantities}
taking into account the biases of the survey and mimicking its algorithms.
A consistent comparison of the galaxy properties is the
only possible way to reliable  test the recipes 
for star formation, feedback and metal enrichment in  
hydrodynamical simulation codes.

We hope this work provides useful resources for
simulators to better compare their galaxies with
SDSS data, and
encourages them to test the effects
of applying observational techniques on the properties
of the simulated galaxies. Making unbiased comparisons
has been proven to be of crucial importance 
to decide on the success or failure of a galaxy formation model, 
offering insights into possible refinements of
galaxy formation codes.


\section*{Acknowledgments}
We thank Yago Ascasibar and Javier Casado G\'omez for useful discussions and 
comments. 
We also acknowledge Michael Aumer for providing his simulations, 
and P.-A. Poulhazan and P. Creasey for sharing the new chemical code.
GG and CS acknowledge support from the Leibniz Gemeinschaft,
through SAW-Project SAW-2012-AIP-5 129,
and from the High Performance Computer in Bavaria (SuperMUC) 
through Project pr94zo.
CJW acknowledges support through the Marie Curie Career Integration Grant 
303912. GG acknowledges support from the DAAD through
the Spain-Germany Collaboration programme (Project ID 57050803).

\bibliographystyle{mn2e}
\bibliography{biblio}

\appendix


\section[]{Stellar age and metallicity determination}
\label{app:stell_age_met}

\begin{figure*}
  \centering
\includegraphics[width=7cm]{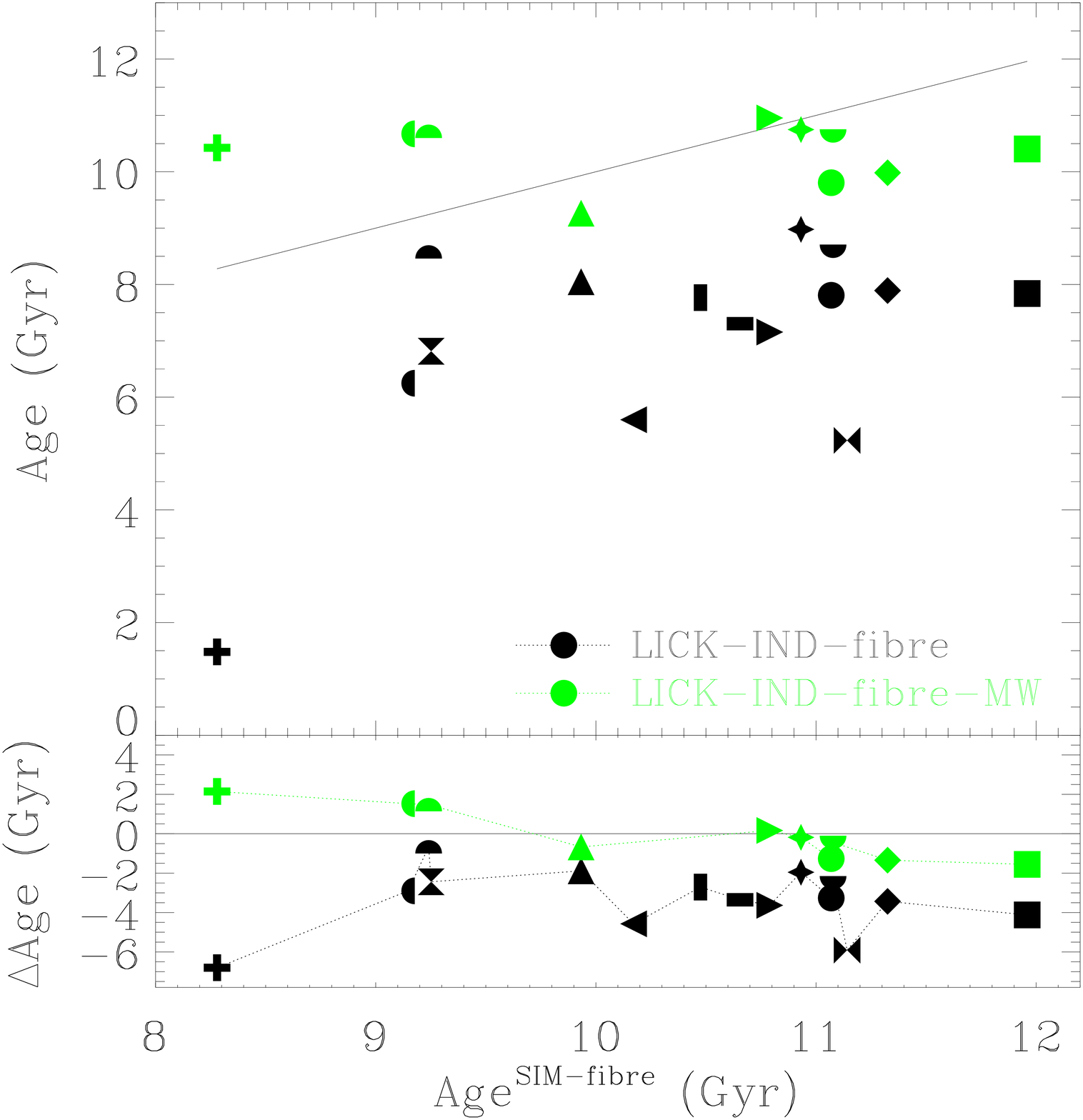}\hspace{0.8cm}\includegraphics[width=7cm]{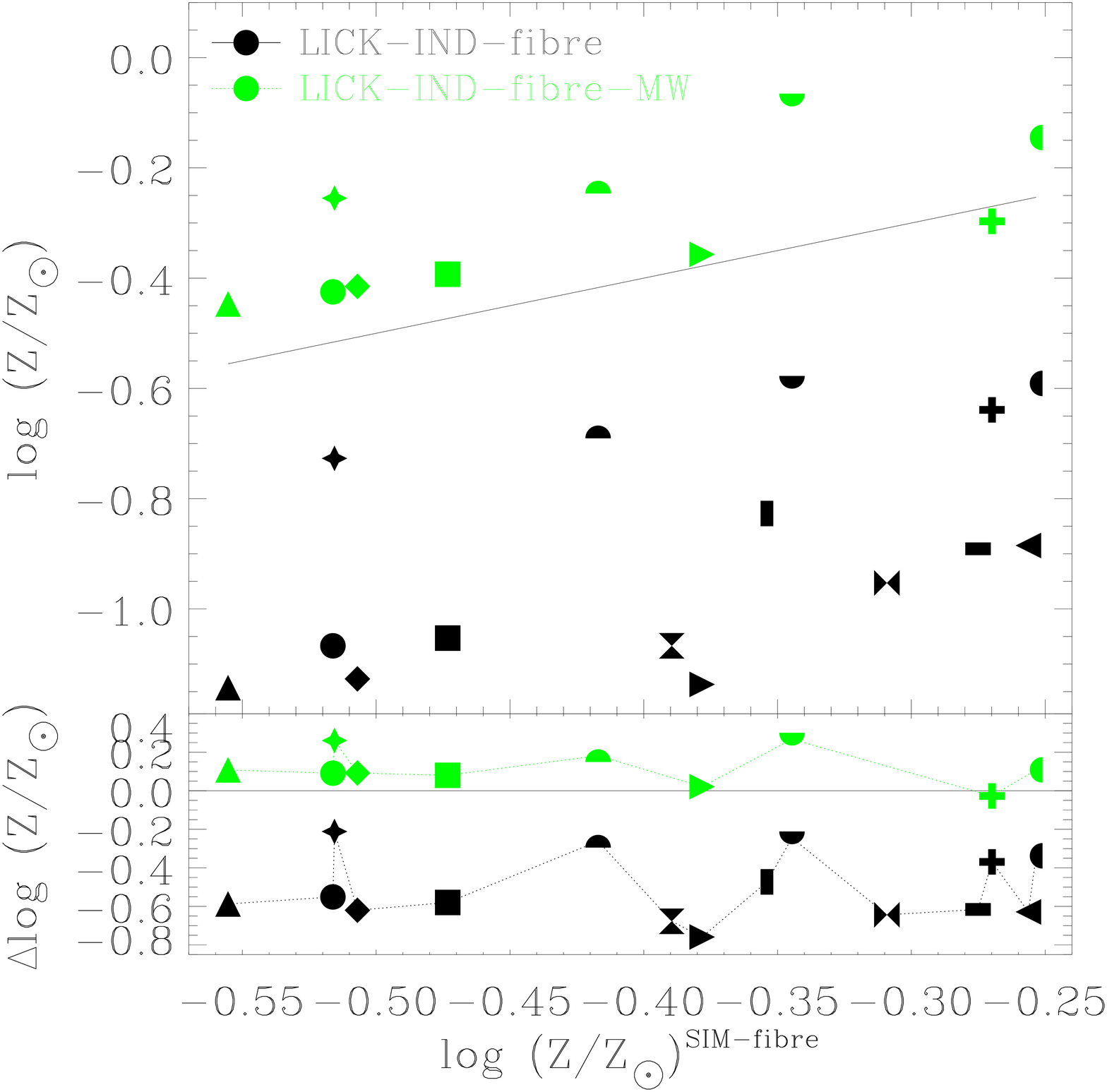}
\includegraphics[width=1.04cm]{figures/magnitudes_label_vertical.eps}
\vspace{0.4cm}
\caption{Mean stellar ages and metallicities calculated 
in the region covered by the SDSS fibre with the 
\citet{Gallazzi05} method, applying two different techniques to extract
the Lick indices from the simulations, in the 1-to-1 relation with the
mean ages/metallicities taken directly from the snapshots. 
The LICK-IND-fibre-MW method has been used for the age and metallicity 
determination for 10 galaxies in \citet{Guidi15}, while the LICK-IND-fibre 
method refers to the calculations done in  this work
for all 15 objects.}
\label{fig:age_met_comparison}
\end{figure*}

In paperI, we have used, among other methods, the
\citet{Gallazzi05}  technique
to determine the mean stellar ages and metallicities of the simulated
galaxies.
As this method is based on fitting selected Lick indices, we 
used the names LICK-IND and LICK-IND-fibre 
to refer, respectively, to the cases
where we consider star particles 
in the full FoV and inside the fibre (Figs. 10-11 in PaperI).
In this work (see
Sec.~\ref{sec:stellar_ages_and_stellar_metallicities}),
we have also used the Gallazzi et al. technique applied to the fibre
spectra (referred to as LICK-IND-fibre
in Fig.~\ref{fig:1_to_1_age}-\ref{fig:1_to_1_met}), 
although
the indices have been extracted from the simulations in a different
way compared to PaperI, as we explain below.

The indices used in PaperI are 
estimated (for 10 galaxies) by deriving the strength of the absorption
features from the BC03 tables, after 
interpolating the values in the tables according to the age and 
metallicity of each stellar particle. The averaged values
of the Lick indices are obtained weighting with the particle mass
and 
are given to the fitting routines.
In contrast, in this work (see description in 
Sec.~\ref{sec:stellar_ages_and_stellar_metallicities}) 
the values of the indices are estimated from
the synthetic spectra generated with {\sc sunrise}, more consistent 
with the Gallazzi et al. method.

In this section we make a comparison between the results
for stellar ages and metallicities obtained using the
two different estimations of the Lick indices. 
In order to avoid confusion with the naming,
in this section we consider only the results in the fibre
and we refer to the calculation done in PaperI 
as LICK-IND-fibre-MW, as
the average values of the indices
have been calculated weighting with the particle 
mass.
The names LICK-IND-fibre are then kept for the estimations
done using the techniques described in this work.

In Fig.~\ref{fig:age_met_comparison} we show the mean stellar 
ages and metallicities obtained 
using the Lick indices method,
as a function of the age/metallicity derived 
directly from the simulations (SIM).
The upper panel of each figure
show the results of the 1:1 relation, while the lower panel shows
the corresponding differences
(note that these figures are similar to Figs. 10-11 of 
PaperI, but only including
the Gallazzi et al. determination).

From the figures it is evident that the determination of a galaxy's stellar age
and metallicity strongly depends on the way the indices are calculated
 in the simulated galaxies; 
for stellar ages (on the left) 
the discrepancies
between the LICK-IND-fibre and LICK-IND-fibre-MW methods 
are significant and increase at younger ages, with the former giving 
systematically younger ages compared to the latter.
 The two methods agree better for ages $\greatsim 10$ Gyr, although the MW 
calculation still predicts galaxies older by $\sim 2-4$ Gyr.
It is important to note that while the MW calculation
predicts galaxies that look
older at younger ages and slightly younger
at older ages ($\greatsim 10$ Gyr) compared to the SIM values,
the calculations used in this work always predict younger galaxies
compared to SIM, without any strong dependence on the mean stellar age.

From the right figure (mean metallicity), we again see
differences between the  metallicity determination obtained with the 
two estimations of the Lick indices.
In particular, the use of the  MW method gives
systematically higher metallicities compared to
LICK-IND,  even by $\sim 1$ dex, without strong dependence on the
metallicity.
Finally, we note that the
MW estimation is in better agreement with the SIM values, 
at least for metal-rich
galaxies, although it predicts in general slightly higher metallicities,
while the 
LICK-IND results exhibit the opposite trend, with systematically lower
metallicities with respect to SIM.

\section[]{Testing the dependence of the observables on SUNRISE free
parameters}
\label{app:sunrise_parameters}

\begin{figure}
  \centering
\includegraphics[width=6.8cm]{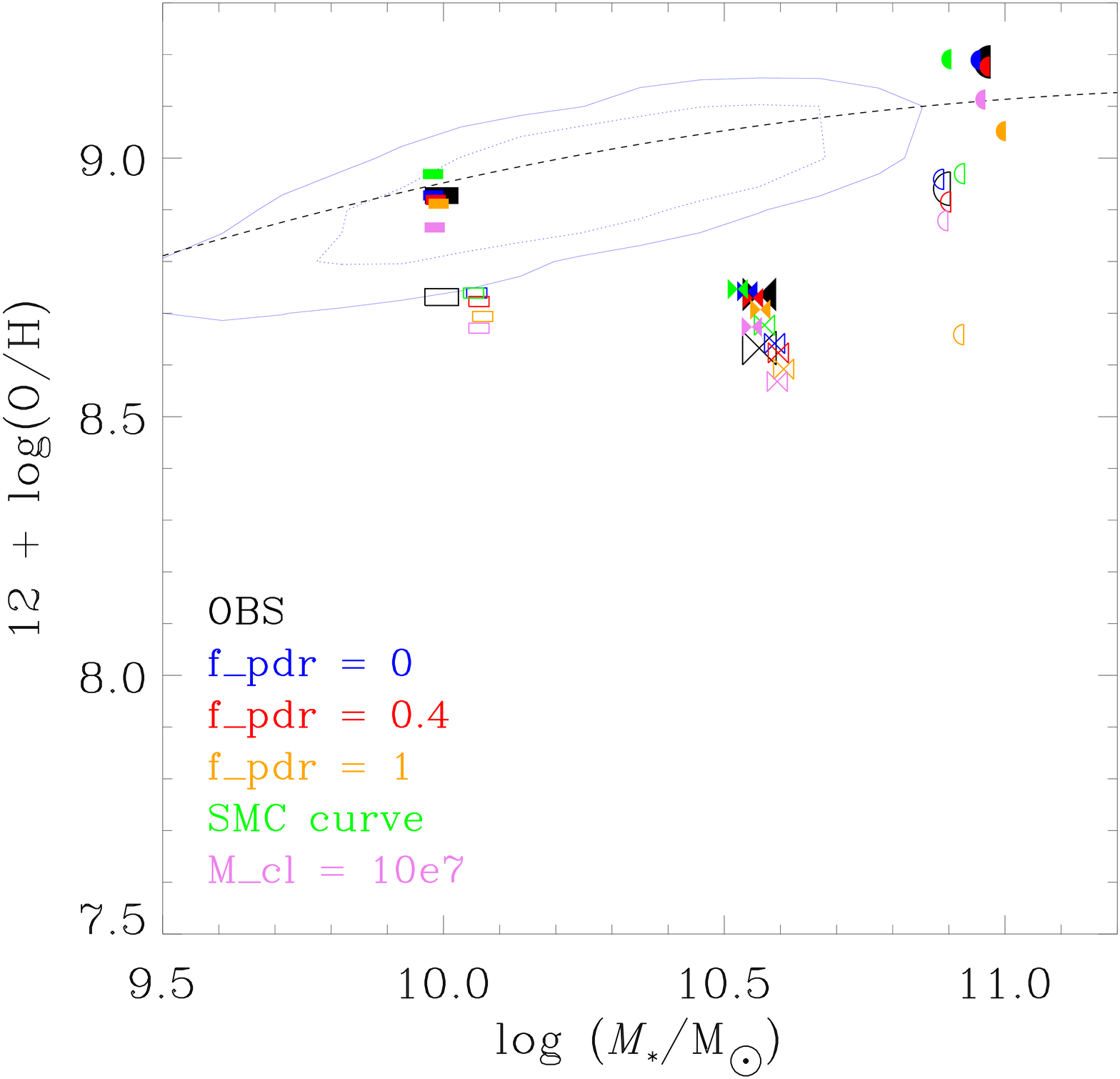}
\caption{Gas metallicity extracted using the T04 calibration applied to 
{\sc sunrise} spectra of 3 simulated galaxies, computed using 
different values for $f_{\text{PDR}}$, 
M$_{cl}$, and a SMC-like dust extinction 
curve. In filled and open symbols we show, respectively, the quantities
computed face-on and edge-on, with the OBS symbols giving the results used in
this work. }
\label{fig:quantities}
\end{figure}

In this section we explore the robustness of our results on the change 
of the {\sc sunrise} free parameters. We have in particular tested the 
dependency of the physical properties
recovered by the observational methods on the choice of the  {\sc mappings} 
parameters $f_{\text{PDR}}$ and M$_{cl}$, and 
on the assumed dust extinction curve in {\sc sunrise}.
We re-run {\sc sunrise} for 3 simulated galaxies, assuming 3 different 
values for $f_{\text{PDR}}$ $\{0, 0.4, 1\}$ and M$_{cl} = 10^7 M_{\odot}$ (while
in this work we have assumed $f_{\text{PDR}} = 0.2$ and M$_{cl} = 10^5 M_{\odot}$). 

We found that the most relevant effects on the magnitude and colour arise by
the use of a SMC dust curve, however with small differences 
$\lesssim 0.05-0.15$ mag in the $r-$band magnitude and $0.05-0.2$ mag
in the $(u-r$) colour. 
The effect on the stellar mass is in most of the cases negligible, lower 
than $\lesssim 0.05$ dex.
The SMC dust curve also affects the concentrations, with differences of 
the order $\lesssim 0.02$, and the S\'ersic index determination 
changing the results by $\lesssim 0.1$, while the effects of the other
parameters is negligible on these two quantities. 
On the contrary, the gas metallicity is the physical quantity more sensible
to the changes in the free parameters, the strongest effects arising by
the choice of $f_{\text{PDR}}$ (especially for $f_{\text{PDR}} = 1$) 
and M$_{cl}$, although in general the differences are $\lesssim 0.1$ dex  
The SFR and sSFR are only slightly affected by the changes in the parameters, 
with offsets of the order $\lesssim 0.1-0.2$ dex.

\bsp

\label{lastpage}

\end{document}